\documentclass{article}

\PassOptionsToPackage{colorlinks=true, citecolor=blue, linkcolor=blue, urlcolor=black}{hyperref}

\usepackage{arxiv}
\usepackage[utf8]{inputenc}
\usepackage[T1]{fontenc}
\usepackage{url}
\usepackage{booktabs}
\usepackage{nicefrac}
\usepackage{microtype}
\usepackage{lipsum}
\usepackage{graphicx}
\usepackage{natbib}
\usepackage{doi}
\usepackage{algorithm}
\usepackage{algorithmicx}
\usepackage{multirow}
\usepackage{multicol}
\usepackage{amsmath,amssymb,amsfonts}
\usepackage{amsthm}
\usepackage{mathrsfs}
\usepackage{xcolor}
\usepackage{float}

\title{A generative foundation model for an all-in-one seismic processing framework}

\author{ \href{https://orcid.org/0000-0001-8868-7967}{\includegraphics[scale=0.06]{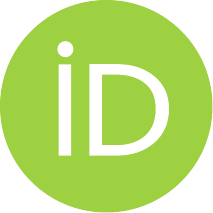}\hspace{1mm}Shijun~Cheng}\\
	Division of Physical Science and Engineering\\
	King Abdullah University of Science and Technology\\
	Thuwal 23955-6900, Saudi Arabia \\
	\texttt{sjcheng.academic@gmail.com} \\
	\And
	\href{https://orcid.org/0000-0002-9616-9761}{\includegraphics[scale=0.06]{orcid.pdf}\hspace{1mm}Randy~Harsuko} \\
	Division of Physical Science and Engineering\\
	King Abdullah University of Science and Technology\\
	Thuwal 23955-6900, Saudi Arabia \\
	\texttt{mochammad.randycaesario@kaust.edu.sa} \\
        \And
	\href{https://orcid.org/0000-0002-9363-9799}{\includegraphics[scale=0.06]{orcid.pdf}\hspace{1mm}Tariq~Alkhalifah} \\
	Division of Physical Science and Engineering\\
	King Abdullah University of Science and Technology\\
	Thuwal 23955-6900, Saudi Arabia \\
	\texttt{tariq.alkhalifah@kaust.edu.sa} \\
}

\renewcommand{\shorttitle}{Generative seismic foundation model}

\hypersetup{
pdftitle={A template for the arxiv style},
pdfsubject={q-bio.NC, q-bio.QM},
pdfauthor={David S.~Hippocampus, Elias D.~Striatum},
pdfkeywords={First keyword, Second keyword, More},
}

\begin{document}
\maketitle

\begin{abstract}
Seismic data often face challenges in their utilization due to noise contamination, incomplete acquisition, and limited low-frequency information, which hinder accurate subsurface imaging and interpretation. Traditional processing methods rely heavily on task-specific designs to address these challenges and fail to account for the variability of data. To address these limitations, we present a generative seismic foundation model (GSFM), a unified framework based on generative diffusion models (GDMs), designed to tackle multi-task seismic processing challenges, including denoising, backscattered noise attenuation, interpolation, and low-frequency extrapolation. GSFM leverages a pre-training stage on synthetic data to capture the features of clean, complete, and broadband seismic data distributions and applies an iterative fine-tuning strategy to adapt the model to field data. By adopting a target-oriented diffusion process prediction, GSFM improves computational efficiency without compromising accuracy. Synthetic data tests demonstrate GSFM surpasses benchmarks with equivalent architectures in all tasks and achieves performance comparable to traditional pre-training strategies, even after their fine-tuning. Also, field data tests suggest that our iterative fine-tuning approach addresses the generalization limitations of conventional pre-training and fine-tuning paradigms, delivering significantly enhanced performance across diverse tasks. Furthermore, GSFM’s inherent probabilistic nature enables effective uncertainty quantification, offering valuable insights into the reliability of processing results.
\end{abstract}

\keywords{Generative foundation model \and Generative diffusion models \and Multi-task seismic processing}
\section{Introduction}
Seismic processing is an essential step for raw data acquisition to produce high-quality subsurface images \citep{yilmaz2001seismic}. It involves a series of complex and diverse procedures aimed at revealing detailed information about subsurface formations and their physical properties. Due to the highly intricate nature of seismic wave propagation in subsurface media and the interference of acquisition environments, raw acquired data are often degraded by various factors. For example, environmental noise reduces the signal-to-noise ratio, making it challenging to extract valuable signals. Damaged geophones can lead to bad traces in the data, compromising the consistency and completeness of subsequent processing. Low-frequency signals are often week, resulting in the loss of crucial signal components that are vital for accurately characterizing subsurface structures \citep{virieux2009overview}. These factors negatively impact the accuracy of subsequent seismic processing, imaging and inversion. Therefore, various seismic processing steps should be performed to enhance data quality, strengthen signals, and eliminate interferences, thereby achieving reliable subsurface imaging results and accurate geological interpretation. 

The conventional seismic processing paradigm generally consists of several key steps designed to address the aforementioned issues and enhance data quality. The first stage is preprocessing, which usually includes denoising to mitigate the impact of noise \citep{abma1995lateral, krohn2008introduction, chen2014random, chen2015random, liu2015signal}. Usually, static correction and normalization are performed to compensate for surface irregularities and variations in amplitude, ensuring consistency in signal phase and amplitude \citep{cox1999static}. Multiple suppression is also often employed to eliminate the impact of multiple reflections, thereby enhancing the clarity of primary reflection signals \citep{verschuur1992adaptive, lopez2015closed}. For areas with incomplete data acquisition, interpolation techniques are used to fill in missing information, thereby improving spatial sampling density and resolution \citep{spitz1991seismic, wang2002seismic, chen2019interpolation}. Moreover, velocity analysis \citep{alkhalifah1995velocity, symes2008migration, fomel2009velocity} and migration \citep{baysal1983reverse, chang1987elastic, zhang2015stable} are central to obtain an image of the subsurface. In this process, accurate velocity models are constructed to reposition seismic reflection events to their true locations, resulting in precise subsurface images \citep{etgen2009overview}. In addition, especially recently, inversion techniques are applied to extract lithological and physical property information from the subsurface, enabling a quantitative description of subsurface formations \citep{tarantola1984inversion, tarantola1986strategy, alkhalifah2014tomography}. These processing steps work together to gradually improve the quality of seismic data, and the resulting image for proper geological interpretation and understanding. 

The advantages of the traditional seismic processing paradigm lie in its rigorous theoretical foundation and its extensive application in geophysical exploration \citep{yilmaz2001seismic}. These processing steps have been validated over time and, also, have demonstrated effectiveness in addressing a variety of complex issues while progressively enhancing the quality and reliability of seismic data. Additionally, traditional methods possess strong physical interpretability, enabling clear imaging of subsurface structures and extraction of crucial lithological and physical properties. However, there are also notable limitations associated with traditional methods. Firstly, conventional seismic processing often relies heavily on expert knowledge and experience, requiring frequent parameter adjustments and expert judgment throughout the various steps, to adapt to the various data. In other words, the processing algorithms, other than certain user-defined parameters, are often fixed and not driven by the data. This results in a high professional threshold and a lengthy processing cycle \citep{yu2021deep}. Secondly, given the increasing volume of data, the efficiency and timeliness of traditional methods struggle to meet practical demands, with the data processing and imaging often consuming considerable time and computational resources \citep{hou2021machine}. Lastly, the performance of traditional methods is often insufficiently robust, making them susceptible to noise and the complexities of subsurface media, which hinders their ability to consistently deliver high-quality results \citep{Li2020DLInversion}. 

To overcome the limitations of traditional methods, neural network (NN)-based seismic processing approaches have gradually gained attention due to their numerous unique advantages \citep{yu2021deep, mousavi2022deep, mousavi2024applications}. For instance, deep learning (DL) methods can automatically learn features from data, thereby reducing the reliance on expert knowledge, while also better meeting the need to handle large volumes of data. Furthermore, NN-based models often exhibit superior performance when processing complex seismic data. Typically, an NN-based seismic processing paradigm involves training a deep NN on a substantial amount of seismic data to approximate the nonlinear relationship between input and target data. Since target data from real-world cases are often inaccessible, a common approach is to train the NN using synthetic data in a supervised learning (SL) manner before applying it to real data \citep{yu2019deep, wang2019deep, dong2019desert, wu2019faultseg3d, wu2020building, zhang2021deep, dong2024can, dong2024seismic}. A significant limitation of this approach arises when the synthetic data distribution poorly represents the real data, leading to considerable performance degradation for the trained network \citep{alkhalifah2022mlreal, zhang2022improving}. Therefore, an alternative method is to use self-supervised learning (SSL) (or unsupervised learning) to eliminate the need for labeled data, enabling the network to be trained directly on real data, which can mitigate the generalization issues \citep{saad2020deep, birnie2021potential, liu2023trace, liu2024self, liu2024gabor, saad2024noise, cheng2024effective, cheng2024self}. However, since the training is performed on each real seismic dataset individually, it is often dataset and task specific and, thus, the overall efficiency is lower compared to networks trained using SL. 


Actually, regardless of whether it is in the SL or SSL paradigm, another major issue is that a trained network is often tailored to a specific seismic processing task (SPT). When switching to another task, the network is often trained again from scratch. As mentioned earlier, seismic processing comprises multiple distinct tasks, and training a network from scratch for each task incurs significant time and the computational cost. Consequently, some recent paradigms based on pre-trained models have been proposed, where these models are first pre-trained on large amounts of seismic data using SSL for reconstruction, and then fine-tuned for downstream tasks to improve training efficiency and reduce computational costs. For example, \cite{harsuko2022storseismic} proposed the StorSeismic framework, in which they pre-trained a Transformer model that takes the sequence of seismic shot gathers as input to extract and store features of the seismic data. The pre-trained model is then fine-tuned for multiple SPTs, such as denoising, velocity estimation, first arrival picking, and normal moveout correction, among other tasks. The fine-tuned model demonstrated excellent performance on field data. Similarly, \cite{sheng2023seismic} introduced the Seismic Foundation Model (SFM), employing the Masked Autoencoders approach to pre-train a Transformer on over 2 million large datasets. After pre-training, they extracted the encoder part and connected a simple decoder network for fine-tuning on downstream tasks. SFM exhibited superior performance across tasks like denoising, interpolation, seismic facies classification, geological body recognition, and inversion. Unlike the previous two paradigms, \cite{cheng2024meta} proposed a Meta-Processing framework for multi-task seismic processing that employs meta-learning to extract shared features of seismic data from very limited datasets, thereby providing a robust initialization. This initialization allows for rapid convergence to optimal performance across various SPTs. 

We can see that, the core of the pre-training strategy lies in leveraging NNs to learn and extract distributional characteristics of seismic data, enabling these pre-trained networks to achieve rapid convergence and outstanding performance across various downstream SPTs. Therefore, it inspires us that if a network model can effectively capture the distribution characteristics of seismic data, it can significantly enhance its performance in seismic processing. Recently, generative diffusion models (GDMs) have shown substantial potential in seismology due to their powerful ability to learn given data distributions, including applications such as denoising \citep{li2024conditional, xiao2024diffusion, trappolini2024cold}, interpolation \citep{wei2023seismic, liu2024generative, wang2024self, wei2024seismic}, resolution enhancement \citep{zhang2024seisresodiff}, waveform separation \citep{zhang2024conditional}, imaging improvement \cite{shi2024generative}, and velocity model building \citep{wang2023prior, wang2024controllable, taufik2024learned}. Notably, \cite{durall2023deep} tested GDMs on various SPTs, including demultiple, denoising, and interpolation. They trained GDMs on synthetic data and evaluated it on synthetic data. They presented results of field data testing for the demultiple task, demonstrating competitive outcomes with traditional DL methods. However, significant signal leakage was still observed, which can be attributed to generalization issues arising from the distributional shift between synthetic and field data. In addition, for different SPTs, they trained different GDMs from scratch to accommodate each specific task, which is time-consuming. 

In this paper, we propose a generative seismic foundation model (GSFM) framework for various SPTs. This framework is based on the GDM's powerful capability to capture and store the distributional characteristics of seismic data, potentially offering greater expressiveness compared to traditional pre-training methods. Due to the GDM's need for target data distributions, as shown by \cite{durall2023deep}, we also train our GDM on synthetic data. However, a significant difference from Durall et al.'s approach is that we train various SPTs simultaneously on a single GDM, such as denoising, backscattered noise attenuation, interpolation, and low-frequency extrapolation. We encode these tasks by introducing different class labels, embedding them into the training process of the GDM, enabling the network to automatically identify and handle various SPTs. Training for multi-task  applications is based on the assumption that ideal seismic data (the target) should be clean, complete, and broadband. By training the GDM to capture this ideal distribution, we enable the model to generate the ideal target output from low-quality seismic data. Additionally, we adopt, within the GDM framework, target prediction instead of noise prediction during training to enhance both training stability and inference efficiency. Predicting the target directly aligns the model output with the ideal seismic data distribution, avoiding the iterative denoising process commonly required in conventional GDMs. This design not only simplifies the training process by reducing optimization complexity but also allows us to achieve high-quality results during inference with just a single sampling step. Nevertheless, due to the feature gap between synthetic and field data, we would still face generalization issues when applying the trained model to field data. To address this problem, we propose a strategy to fine-tune our pre-trained GDM on field data using an SSL approach. Specifically, during the initial stage of fine-tuning for each SPT, we use the pre-trained GDM model to directly predict the field data, which is then added to our training dataset. After the GDM model undergoes several iterations of optimization, we iteratively employ the model trained in the previous stage to predict field data and update the training set at fixed intervals. In this way, we gradually shift the distribution captured by the pre-trained GDM from the synthetic domain to the field data distribution, thereby enhancing the model's performance on field data. 

Furthermore, the inherent randomness in the initial noise used during the sampling process allows us to generate multiple predictions for the same input condition. This provides a natural mechanism to assess prediction variability. By evaluating the standard deviation of these predictions, we can identify regions of higher uncertainty, which often correspond to areas with greater signal leakage or processing errors. This capability not only helps evaluate the reliability of the model's predictions but also provides valuable feedback to guide further optimization during the fine-tuning process, ensuring robust performance on field data. 

The contributions of this paper can be summarized as follows:
\begin{itemize}
   \item We propose a generative seismic foundation model framework capable of simultaneously performing various SPTs.
   \item We introduce the use of class label constraints to guide the NN in jointly optimizing different SPTs.
   \item We propose a strategy to fine-tune the pre-trained foundation model on synthetic data using an SSL approach on field data, thereby overcoming the generalization issues of NNs.
   \item We leverage the probabilistic nature of GDMs to quantify the uncertainty of the processing product, which helps assess its reliability and helps guide the fine-tuning of the pre-trained model.
   \item Examples from synthetic and field data demonstrate that our all-in-one seismic processing framework can achieve good processing performance.
\end{itemize}

\section{Review of conventional neural network-based seismic processing}
Traditional seismic processing methods often rely on explicit physical models and assumptions, which may not be fully applicable in complex media. In contrast, the advantage of NNs lies in their ability to automatically approximate such complex mapping relationships by extracting features from data, without relying heavily on prior assumptions.

Commonly, NN-based seismic processing methods can be viewed as a parameterized function approximator, which adjusts its internal weights through a training process to learn the nonlinear relationship that maps seismic data, $x_i$, (such as raw noisy data) to the desired output products, $y_i$, (such as denoised data). During this process, the network optimizes its parameters by minimizing the error between the NN output (prediction) and the target, $y_i$, to capture key features in the seismic data, which can be represented by the following loss function:
\begin{equation}\label{eq1}
L(\theta) = \frac{1}{N} \sum_{i=1}^N \| \text{NN}_\theta(x_i) - y_i \|^2,
\end{equation}
where $\theta$ is the set of parameters of the network, $\text{NN}_\theta(x_i)$ is the predicted output of the network for input $x_i$, $y_i$ is the corresponding target output, and $N$ is the number of training samples. By minimizing the loss function $L(\theta)$, the network continuously optimizes its parameters to make the predicted results as close as possible to the target output. 

To further enhance the performance of conventional NN-based seismic processing, a pre-training and fine-tuning paradigm has been proposed \citep{harsuko2022storseismic, sheng2023seismic}. The pre-training method involves an SSL training by reconstructing masked original seismic data, thus providing a good initial parameter set for downstream tasks. The objective can be expressed as the following loss function minimization problem:
\begin{equation}\label{eq2}
L(\theta) = \frac{1}{N} \sum_{i=1}^N \| \text{NN}_\theta(x_i^{masked}) - x_i \|^2
\end{equation}
where $\text{NN}_\theta(x_i^{masked})$ is the NN's reconstructed output for the masked input $x_i^{masked}$, and $x_i$ is the corresponding original input data. By performing the pre-training phase, the network can learn representations of the basic features of the original seismic data. On this basis, fine-tuning training on labeled datasets allows the network to better adapt to specific task requirements. This paradigm not only significantly improves the generalization performance of the network but also accelerates convergence and reduces the dependence on labeled data, thereby achieving more robust seismic processing under complex conditions. 

Despite the notable progress made by the current pre-training and fine-tuning paradigm in seismic data processing, we still face several challenges. First, the pre-training stage often relies on synthetic data, resulting in limited generalizability when addressing the complexities of real data. Meanwhile, due to the differences between pre-training tasks (e.g., reconstruction) and downstream tasks (e.g., denoising, interpolation, or low-frequency extrapolation), the model's performance may be constrained during task transfer, preventing it from fully leveraging the benefits of pre-training phase. Moreover, the dependency on labeled data during fine-tuning stage further restricts the paradigm’s applicability when labeled data are scarce. 

To address these challenges, we, in the following, present a framework for a generative seismic foundation model (GSFM) based on generative diffusion models (GDM). This framework aims to capture ideal seismic data distribution features through multi-task learning and generative data distribution modeling, while training on synthetic data for various tasks to enable the model to handle various seismic processing tasks (SPTs) effectively. By incorporating task-specific encoding, the model can automatically identify and manage different SPTs during training. Furthermore, we introduce a gradual transfer strategy using an SSL approach to fine-tune the model on real data, progressively shifting its distribution from synthetic data to real data, thereby improving its performance in practical applications. 
\section{Method}
In this section, we will first introduce the fundamental concepts of GDMs. Following that, we will present the framework for a GSFM based on GDMs. We will provide detailed illustrations on how we perform multi-task encoding, pre-training, fine-tuning, and prediction. Finally, we will introduce our network architecture. 

\subsection{Generative diffusion models}
GDMs are gaining attention for its strong capability to produce highly realistic samples. These models initially convert data into pure noise through a forward process and then progressively denoise it to recover the data in the reverse process. 

Within the denoising diffusion probabilistic model (DDPM), the forward process is defined as \citep{ho2020denoising}:
\begin{equation}\label{eq3}
q(x_t | x_{t-1}) = \mathcal{N}(x_t; \sqrt{\alpha_t} x_{t-1}, (1 - \alpha_t) \mathbf{I}).
\end{equation}
In the forward process, the data sample \( x_0 \) gradually transforms into a noisy version \( x_t \) at each step \( t \), controlled by the parameter \( \alpha_t \), with \( \alpha_t = 1 - \beta_t \), where \( \beta_t \) is a small constant specifying the noise variance incrementally introduced at each step. Thus, at each step, noise is added according to the Gaussian distribution \( \mathcal{N} \).  This process adds isotropic Gaussian noise, facilitated by the identity matrix \( \mathbf{I} \), and progressively diffuses the original data into a Gaussian noise distribution from step \( t = 1 \) to $t = T$. 

To express \( x_t \) directly in terms of \( x_0 \) and a noise term, we can use the reparameterization trick:
\begin{equation}\label{eq4}
x_t = \sqrt{\bar{\alpha}_t} x_0 + \sqrt{1 - \bar{\alpha}_t} \epsilon,
\end{equation}
where \( \epsilon \sim \mathcal{N}(0, \mathbf{I}) \) is Gaussian noise, and \( \bar{\alpha}_t = \prod_{i=1}^t (1 - \beta_i) \). 

To learn how to reverse this process, the network is trained to predict the noise \( \epsilon \) added to the sample at each step \( t \). Given a noisy sample \( x_t \), the network predicts the noise term \( \epsilon_\theta(x_t, t) \). The objective of the network is to minimize the difference between the actual noise \( \epsilon \) (added during the forward process) and the predicted noise \( \epsilon_\theta(x_t, t) \). This is achieved by minimizing the mean squared error (MSE) between the added noise and the predicted noise:
\begin{equation}\label{eq5}
L(\theta) = \mathbb{E}_{x_0, \epsilon, t} \left[ \| \epsilon - \epsilon_\theta(x_t, t) \|^2 \right]
\end{equation}
This objective trains the network to accurately predict the noise added at each step, and to do so, the network needs to store the features of the signal information diffused by this added noise. 

Once the network is trained, the reverse process, starting for a noise sample drawn from the Gaussian distribution, uses the network’s noise predictions \( \epsilon_\theta(x_t, t) \) to iteratively inject the stored signal and gradually remove noise to reconstruct a sample from the distribution it was trained on. Specifically, starting from pure noise \( x_T \), each step in the reverse process estimates the previous sample \( x_{t-1} \) from \( x_t \) by subtracting the predicted noise component. The reverse sampling equation can be expressed as:
\begin{equation}\label{eq6}
x_{t-1} = \frac{1}{\sqrt{\alpha_t}} \left( x_t - \frac{1 - \alpha_t}{\sqrt{1 - \bar{\alpha}_t}} \epsilon_\theta(x_t, t) \right) + \sigma_t z,
\end{equation}
where \( \epsilon_\theta(x_t, t) \) is the noise predicted by the trained network, \( \sigma_t z \) is an optional noise term for stochastic sampling, and \( z \sim \mathcal{N}(0, \mathbf{I}) \). 

In DDPM, the step-by-step denoising process is implemented through a Markov chain, which requires numerous time steps to gradually remove noise. As a result, the sampling speed of DDPM is relatively slow. In contrast, the denoising diffusion implicit model (DDIM) improves the sampling process of DDPM by removing the dependence on the Markov chain \citep{song2020denoising}. Actually, DDIM and DDPM share the same forward process, which can also control noise introduction through Equation \ref{eq4}. The reverse process of DDIM can be represented by the following sampling equation:
\begin{equation}\label{eq7}
x_{t-1} = \sqrt{\bar{\alpha}_{t-1}} \hat{x}_0 + \sqrt{1 - \bar{\alpha}_{t-1} - \sigma_t^2} \, \epsilon_\theta(x_t, t) + \sigma_t z,
\end{equation}
with
\begin{equation}\label{eq8}
\hat{x}_0 = \frac{x_t - \sqrt{1 - \bar{\alpha}_t} \, \epsilon_\theta(x_t, t)}{\sqrt{\bar{\alpha}_t}} 
\end{equation}
giving an estimate of the original data \( x_0 \) directly, \( \epsilon_\theta(x_t, t) \) is the noise term predicted by the neural network, and \( z \sim \mathcal{N}(0, \mathbf{I}) \) is a random noise term, and \(\sigma_t\) controls the level of this randomness. When \(\sigma_t = 0\), the sampling process becomes deterministic because the random noise term \( \sigma_t z \) is removed. If \(\sigma_t \neq 0\), it introduces a small random term \( \sigma_t z \) in the sampling, allowing for some variation in the generated samples. Since we aim to apply DDIM in seismic processing, we prefer a deterministic approach rather than producing random solutions. Therefore, we will set parameter \(\sigma_t = 0\) in the following.

In DDIM, an additional improvement involves directly training the network to predict the original clean image $x_0$ rather than focusing on the noise $\epsilon$ added to $x_0$ \citep{bansal2024cold}. This approach benefits from effectively leveraging the spatial coherence and semantic information within the image, enabling faster convergence and higher generation quality. In this case, the network's optimization target shifts to minimizing the difference between the predicted image $x_{0,\theta}(x_t, t)$ and the original image $x_0$, as follows:
\begin{equation}\label{eq9}
L(\theta) = \mathbb{E}_{x_0, \epsilon, t} \left[ \| x_0 - x_{0,\theta}(x_t, t) \|^2 \right]. 
\end{equation}

In the $x_0$-based prediction framework, the reverse sampling equation in DDIM can be simplified to:
\begin{equation}\label{eq10}
x_{t-1} = \sqrt{\bar{\alpha}_{t-1}} x_{0,\theta}(x_t, t) + \sqrt{1 - \bar{\alpha}_{t-1}} \hat{\epsilon}(x_t, t)
\end{equation}
where $\hat{\epsilon}(x_t, t)$ is an estimate of the added noise. Given Equation \ref{eq8} in the forward process, we can estimate the added noise $\hat{\epsilon}(x_t, t)$ in terms of $x_t$ and the network's prediction $x_{0,\theta}(x_t, t)$, as follows:
\begin{equation}\label{eq11}
\hat{\epsilon}(x_t, t) = \frac{x_t - \sqrt{\bar{\alpha}_t} x_{0,\theta}(x_t, t)}{\sqrt{1 - \bar{\alpha}_t}}.
\end{equation}

GDMs demonstrate excellent performance in generating high-quality samples due to its strong capability in capturing distributions and its stepwise denoising approach \citep{rombach2022high}. Specifically, DDIM significantly enhances sampling speed by eliminating the dependency on the Markov chain. Furthermore, by adopting a strategy based on predicting $x_0$, we are able to further improve both generation quality and sampling speed, which is needed to meet the accuracy and efficiency requirements of seismic processing. Based on this, we introduce the GSFM framework in the following section, which incorporates multi-task learning into GDM to handle a variety of SPTs, including denoising, interpolation, and low-frequency extrapolation in a unified manner.

\subsection{Generative seismic foundation model: Pre-training}
Our GSFM is adapted from a GDM and employs multi-task simultaneous pre-training on synthetic data, followed by direct fine-tuning on real data. However, in traditional GDMs, the model's input consists of a noisy version of a clean single-channel image, which is used to train the model for stepwise denoising. 

To accommodate the needs of multi-task seismic processing, we extend the input of our GSFM to a dual-channel structure. During the pre-training and fine-tuning phases, the dual-channel inputs may contain different content. In this section, we first explain how the dual-channel network inputs are configured during the pre-training phase. Since the network is optimized on synthetic data during the pre-training phase, we can access the labels for different tasks. Therefore, in this phase, the first channel contains a noisy version of the labels (the target complete clean data), while the second channel is used for the corresponding data to be processed, i.e., the degraded data specific to the task. The content of the second channel varies depending on the SPT, enabling the model to adapt flexibly to different tasks based on the input data. Specifically, the dual channels for different specified tasks are as follows:
\begin{itemize}
    \item \textbf{Denoising}: The second channel contains the data contaminated with noise we want the network to learn to remove.
    \item \textbf{Backscattered noise attenuation}: As a special case, the second channel contains data contaminated with backscattered noise. 
    \item \textbf{Interpolation}: The second channel contains data with missing traces. 
    \item \textbf{Low-frequency extrapolation}: The second channel contains data lacking low-frequency components. 
\end{itemize}

In the pre-training phase, the forward process of our GSFM shares the noise injection formulation of the conventional diffusion model, as shown in Equation \ref{eq4}, thereby constructing the content for the first channel of the dual-channel input. For the second channel, we can see that we essentially use the same input data as that used in conventional NN-based seismic processing methods. 

To enable simultaneous training for different tasks, we introduce a task encoding label $c$, allowing the network to identify and distinguish between various SPTs. For the tasks considered in this paper, including denoising, backscattered noise attenuation, interpolation, and low-frequency extrapolation, their class labels $c$ are defined as 0, 1, 2, and 3, respectively. The embedding method for the task encoding label $c$ is similar to that used for the step $t$. In the section \ref{network_architecture}, Network architecture, we will detail the specific embedding implementation for the task encoding label $c$. 

As previously mentioned, setting the GDM network's prediction target to $x_0$ can enhance generation quality and efficiency. Therefore, during the pre-training and also the following fine-tuning phase, the prediction target is set to $x_0$. In other words, for different SPTs, the network's prediction target corresponds to their respective labeled data. In this case, our pre-training objective can be expressed as:
\begin{equation}\label{eq12}
L(\theta) = \mathbb{E}_{x_0, x, \epsilon, t, c} \left[ \| x_0 - x_{0,\theta}(x_t, x, t, c) \|^2 \right], 
\end{equation}
where $x$ represents the second channel input serving as the conditional constraint. 



Here, we consider the four SPTs described above. However, we emphasize that our framework is flexible and can be extended to accommodate additional SPTs by simply defining the appropriate degraded data format for the second channel and assigning a new class encoding label for each added task. This adaptability allows our GSFM to serve as a versatile foundation for a wide range of seismic processing needs. For example, if our objective is for GSFM to remove surface multiples, we simulate shot gathers with free surface boundary condition to serve as input to the second channel, while having our clean data target modeled using absorbing boundary condition \citep{harsuko2024optimizing}.

\subsection{Generative seismic foundation model: Fine-tuning}\label{finetune_section}
After completing pre-training on synthetic data, our GSFM is directly fine-tuned on real data to enhance its generalization capability for practical applications. During the fine-tuning phase, due to the lack of labels, we employ an SSL-based optimization approach, maintaining the model's adaptability and stability across multi-task seismic processing. To ensure consistency, the fine-tuning process retains the embedding methods for the task encoding label $c$ during pre-training, enabling the model to continue supporting multi-task learning on real data and improving task transfer efficiency. The prediction target during fine-tuning remains set to $x_0$ (pseudo-labels), which represents the ideal output for each task. 

To accomplish fine-tuning, we perform this process independently for each SPT, using the pre-trained network as the starting point for each task and setting the task encoding label $c$ to the value corresponding to the desired task. We propose the following three fine-tuning strategies for each SPT:
\begin{itemize}
    \item \textbf{1}. The pre-trained model on synthetic data is used directly on the raw field data to generate preliminary processing products, which are then used as pseudo-labels during the fine-tuning phase. In this case, the first channel of the network input is a noisy version of the predicted pseudo-labels, while the second channel takes in shot gathers from the field data.

    \item \textbf{2}. The second channel of the network input differs from that of strategy 1. Instead of using the field data, we use a corrupted version (similar to the corruptions applied to the synthetic data) of the pseudo-labels. For example, for the denoising task, additional noise is added to the predicted pseudo-labels. For the backscattered noise attenuation task, backscattered noise is added to the pseudo-labels. For the interpolation task, traces are removed from the pseudo-labels. For the low-frequency extrapolation task, the low-frequency components are filtered out from the pseudo-labels.

    \item \textbf{3}. The third strategy is based on strategy 2 and involves iteratively updating the training dataset during the fine-tuning process. 
    The complete workflow is detailed in Algorithm \ref{alg1}. Specifically, the fine-tuning process is divided into multiple stages, with each stage consisting of several iterations. In the first stage, we maintain the configuration of strategy 2. In each subsequent stage, the model fine-tuned from the previous stage is used directly on the field data, generating new pseudo-labels. Diffusion process is added to these pseudo-labels to create the input for the first channel, while the second channel contains a further corrupted version of the newly generated pseudo-labels, consistent with strategy 2. 
\end{itemize} 

In subsequent experiments, we will test these three fine-tuning strategies to determine which one performs better in enhancing the model's generalization ability and processing performance. Based on our test results, the fine-tuned network induced by strategy 3 provides superior performance. This outcome is expected, as the multi-stage strategy with gradual optimization allows the model to achieve a smooth transition between the feature distributions of synthetic and real data. By updating and further degrading the pseudo-labels at each stage, the model progressively shifts from the synthetic domain to the real data domain during the fine-tuning process. This stepwise adjustment not only makes the model more robust in handling the complexity of real data but also effectively reduces the distribution gap between synthetic and real data, thereby improving the model's generalization capability in real-world tasks.

\begin{algorithm}
\caption{Iterative Fine-Tuning with Progressive Pseudo-Labeling for GSFM}\label{alg1}
\textbf{Input:} Pre-trained GSFM model \\
\textbf{Input:} Raw field data $x$, initial pseudo-labels $x_{\text{pseudo}}$, noise $\epsilon$ \\
\textbf{Input:} Total stages $S$, iterations per stage $N_{\text{stage}}$ \\
\textbf{Input:} Task-specific corruption $\text{COR}[\cdot]$ \\
\textbf{Output:} Fine-tuned GSFM model \\
\textbf{-------------------------------- Fine-Tuning Process -----------------------------} 
\begin{algorithmic}
\State 1: Load pre-trained GSFM model.
\State 2: Set task-specific label $c$ according to the seismic processing task.
\State 3: Initialize pseudo-labels $x_{\text{pseudo}}$ by predicting on field data $x$ with the pre-trained model.
\State 4: \textbf{for} {stage $s = 1$ to $S$} \textbf{do}
\State 5: \quad \quad \textbf{for} {iteration $n = 1$ to $N_{\text{stage}}$} \textbf{do}
\State 6: \quad \quad \quad \quad Sample a step $t$ and add noise to pseudo-labels $x_{\text{pseudo}}$: $x_t = \sqrt{\bar{\alpha}_t} x_{\text{pseudo}} + \sqrt{1 - \bar{\alpha}_t} \epsilon$
\State 7: \quad \quad \quad \quad Apply task-specific corruption to $x_{\text{pseudo}}$: $\hat{x} = \text{COR}[(x_{\text{pseudo}}, c)]$.
\State 8: \quad \quad \quad \quad Forward pass $(x_t, \hat{x}, t, c)$ through the model: $x_{0,\theta}(x_t, \hat{x}, t, c) = \text{GSFM}(x_t, \hat{x}, t, c) $
\State 9: \quad \quad \quad \quad Compute loss with respect to target $x_{\text{pseudo}}$ for this task: \\ 
\quad \quad \quad \quad \quad \quad \quad \quad $L(\theta) = \mathbb{E} \left[ \| x_{\text{pseudo}} - x_{0,\theta}(x_t, \hat{x}, t, c) \|^2 \right]$
\State 10: \quad \quad \quad \quad Backpropagate the loss and update model parameters $\theta$
\State 11: \quad \quad \textbf{end for}
\State 12: \quad \quad After $N_{\text{stage}}$ iterations, use the fine-tuned model to generate the updated pseudo-labels: \\
\quad \quad \quad \quad \quad \quad \quad \quad $x_{\text{pseudo}} = \text{GSFM}(\epsilon, x, t, c)$
\State 13: \textbf{end for}
\State 14:  \textbf{Return:} Fine-tuned GSFM model
\end{algorithmic}
\end{algorithm}

\subsection{Generative seismic foundation model: Predicting}

After completing the fine-tuning process, our GSFM is ready to perform predictions for each seismic processing task independently. For each task-specific fine-tuned network, we obtain predictions tailored to the corresponding task. Unlike conventional NN-based seismic processing methods, which typically use a direct mapping approach, our GSFM leverages a generative prediction process due to its foundation in GDM. 

Specifically, for each SPT, we begin by assigning the corresponding task encoding label $c$ to indicate the target task. The network input is structured as follows: The first channel is initialized with random noise $\epsilon$, while the second channel contains seismic data $x$ that needs processing. Using the reverse process of GDM, we iteratively denoise the input to generate the desired output $x_0$. At each step $t$ in the reverse process, the model estimates $x_0$ based on the current noisy input $x_t$, and the reverse step is given by:
\begin{equation}\label{eq13}
x_{t-1} = \sqrt{\bar{\alpha}_{t-1}} \, x_{0,\theta}(x_t, x, t, c) + \sqrt{1 - \bar{\alpha}_{t-1}} \, \hat{\epsilon}(x_t, x, t, c),
\end{equation}
where $x_{0,\theta}(x_t, x, t, c)$ is the model's prediction of the clean data $x_0$ given the noisy input $x_t$ at step $t$ and task label $c$. Here, $\hat{\epsilon}(x_t, x, t, c)$ represents an estimate of the noise component, which can be computed as in Equation \ref{eq11}. 

The conventional prediction process continues iteratively, with the model starting from a high level of noise and gradually refining the input. The final output at last step, $x_0$, represents the processed data for the specified task, having been transformed from noise to the desired form through a series of denoising steps. However, considering the efficiency requirements in actual processing, we will only use one time step for the sampling process here. Specifically, we only use the last sampling step, that is, $t$ is set to 0 to get our final prediction product. \\

\subsection{Network architecture}\label{network_architecture}

\begin{figure*}[!t]
\centering
\includegraphics[width=\textwidth]{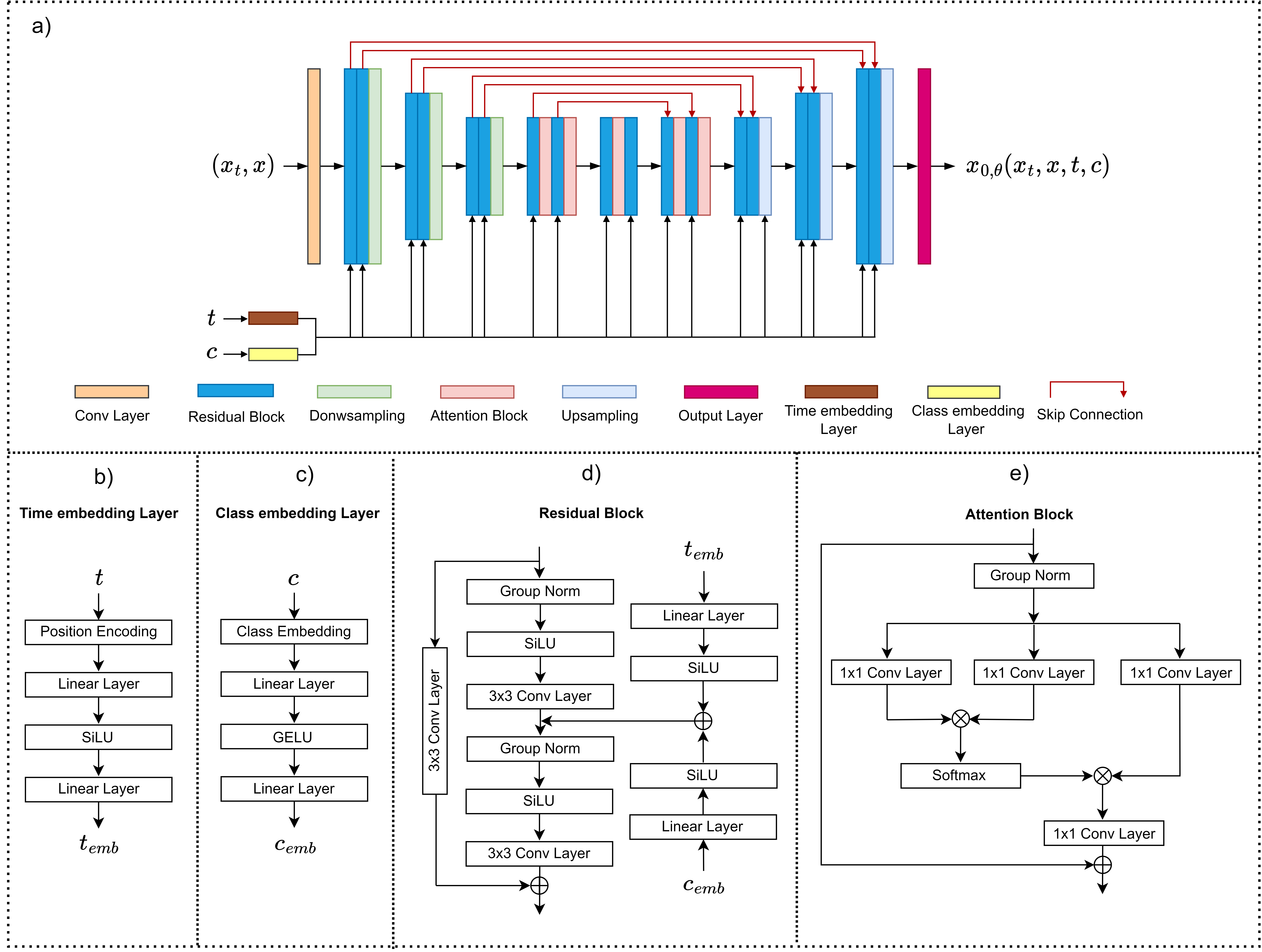}
\caption{An illustration of our network architecture. (a) The overall network structure. (b) Time embedding layer. (c) Class embedding layer. (e) The residual block. (d) The attention block. }
\label{fig1}
\end{figure*}

Our GSFM adopts an enhanced U-Net-based architecture tailored for multi-task seismic processing. This architecture incorporates multi-scale feature extraction, task-specific embeddings, and attention mechanisms to deliver accurate and robust predictions. The main components of the network are illustrated in Figure \ref{fig1}, including convolutional layers, residual blocks, attention blocks, downsampling and upsampling layers, and embeddings for time and task-specific information. 

The GSFM processes dual-channel inputs $(x_t, x)$, where $x_t$ represents in training the target data input at timestep $t$, and $x$ contains the data to be processed specific to the task. These inputs are first passed through an initial $3 \times 3$ convolutional layer that maps the two input channels to 64 feature channels, preparing the data for hierarchical processing in the encoder-decoder structure. 

The encoder path progressively extracts hierarchical features using a combination of downsampling layers, residual blocks and attention blocks. Each downsampling layer reduces the spatial resolution by a factor of 2 and simultaneously doubles the number of feature channels, enabling the extraction of high-level features at coarser scales. Specifically, after the first downsampling operation, the number of channels increases from 64 to 128. Subsequent downsampling operations further increase the channels to 256 and 512. 

The decoder path restores the spatial resolution and reduces the number of channels in a symmetrical manner with respect to the encoder, combining high-level semantic information from the encoder with low-level spatial details via skip connections. Each upsampling layer doubles the spatial resolution and halves the number of feature channels. For example, the number of channels decreases from 512 to 256 after the first upsampling layer. This process continues until the final layer restores the original spatial resolution and reduces the channels back to 64. At the end of the decoder, a final output layer is applied. This layer consists of group normalization, followed by a sigmoid linear unit (SiLU) activation function, and a $3 \times 3$ convolutional layer that reduces the feature channels to the single-channel prediction $x_{0,\theta}(x_t, y, t, c)$. 

To address the requirements of multi-task processing, GSFM integrates two types of embeddings to guide the network with temporal and task-specific information:
\begin{itemize}
    \item \textbf{Time embedding layer} (Figure \ref{fig1}b): The timestep \(t\) is encoded using a sinusoidal positional encoding scheme \citep{vaswani2017attention}, which represents temporal information as a combination of sine and cosine functions. The resulting encoded vector is passed through a series of linear transformations and SiLU activation functions, producing the time embedding vector $t_{emb}$. This embedding vector is injected into the residual blocks to regulate the denoising process across timesteps.
    \item \textbf{Class embedding layer} (Figure \ref{fig1}c): Task-specific information is provided through a learnable embedding layer implemented using \texttt{torch.nn.Embedding}. The task encoding label $c$ is mapped to a high-dimensional embedding vector, which is further processed by linear transformations and gaussian error linear unit (GELU) activations, producing the class embedding vector $c_{emb}$. This embedding vector is incorporated into residual blocks to enable task-specific adaptability.
\end{itemize}

In our GSFM, feature extraction and refinement rely on the integration of residual and attention blocks:
\begin{itemize}
    \item \textbf{Residual blocks} (Figure \ref{fig1}d): Each residual block processes feature maps using a combination of group normalization, SiLU activation functions, and $3 \times 3$ convolutional layers. Task and time embeddings ($c_{emb}$ and $t_{emb}$) are incorporated by projecting them through linear layers and adding the resulting vectors to the feature maps. This design enables task- and time-aware feature processing.
    \item \textbf{Attention block} (Figure \ref{fig1}e): Attention blocks, which is developed by \cite{vaswani2017attention}, are applied to refine the feature maps further. These blocks compute query, key, and value matrices via $1 \times 1$ convolutional layers and normalize attention scores using a softmax operation. The resulting weighted feature maps are aggregated and processed through another $1 \times 1$ convolutional layer. This mechanism allows the network to focus on task-relevant regions, improving feature representation for seismic data.
\end{itemize}

The GSFM leverages downsampling and upsampling layers to capture features across multiple spatial scales. Downsampling layers reduce the spatial resolution of feature maps, enabling the extraction of high-level semantic features, while upsampling layers restore spatial resolution to match the input dimensions. Skip connections link encoder and decoder layers, combining fine-grained spatial details with deep semantic features, thereby improving the accuracy of task-specific predictions.

\section{Synthetic data examples}
In this section, we first introduce the pre-training details of the GSFM, including dataset preparation and training configuration. We then evaluate the pre-trained GSFM's performance on denoising, backscattered noise attenuation, interpolation, and low-frequency extrapolation tasks using synthetic test data. 

To assess the effectiveness of the pre-trained GSFM, we provide two comparative experimental benchmarks:
\begin{itemize}
    \item \textbf{Benchmark 1}: Traditional NN-based processing paradigm. \\
     This benchmark utilizes conventional NN-based seismic processing methods, employing the networks to approximate the nonlinear relationship between input data and target data. To ensure a fair comparison, Benchmark 1 adopts the same U-Net-based architecture as GSFM, but excludes the time encoding module used in the diffusion model. The network takes single-channel degraded data as input and outputs the corresponding target data. This comparative benchmark effectively demonstrates the advantages of a diffusion model-guided network over traditional training approaches.

     \item \textbf{Benchmark 2}: Conventional pre-training and fine-tuning strategy. \\
     In this benchmark, we first pre-train a NN on synthetic data using an SSL approach, followed by fine-tuning for denoising, backscattered noise attenuation, and low-frequency extrapolation tasks. Again, to ensure fairness, we use the same U-Net architecture as GSFM but remove both the task encoding and time encoding modules. During the pre-training phase, we use the GSFM dataset for all tasks, constructing the input data using random masking. During fine-tuning phase, the training is conducted on each individual task's dataset to evaluate performance across different SPTs.
\end{itemize}

\subsection{Pre-training configuration}
Creating synthetic subsurface models that represent the real Earth remains a challenge. For our purposes, we closely follow the workflow introduced by \cite{ovcharenko2022multi} to generate random velocity models, which have been shown to effectively generalize to real data. Specifically, first, we randomly create 1D compressional wave velocity ($V_p$) profiles using velocity values within our expected range of 1,500 to 4,500 m/s. These 1D profiles are then spread laterally to build 2D laterally homogeneous layered velocity models. Lastly, we apply random elastic transforms to the velocity models to distort them and introduce structures resembling realistic geological phenomena (folding, intrusion, etc.) 

Since we aim to establish a foundation model applied to seismic waveforms, it is of utmost importance that the synthetic waveform for the training dataset is as close as a realistic waveform, which justifies the need to use an elastic modeling engine. We use a Pytorch-based seismic modeling and inversion package called Deepwave \citep{richardson_alan_2023} to perform 2D elastic forward modeling on the aforementioned velocity models. The shear wave velocity ($V_s$) is obtained through a fixed ratio of $V_p /\sqrt(3)$, while the density ($\rho$) is obtained through Gardner's relation \citep{gardner1974formation}. The discretization of the subsurface parameters for the modeling is detailed in Table \ref{tab1}. For the acquisition setting, we consider a marine environment where the data is acquired through a towed streamer consisting of an array of hydrophones that records a pressure component from an airgun source. The airgun source is represented by a Ricker wavelet with a peak frequency of 7 Hz. We set the number of receivers to 324 for every shot. More details of the acquisition parameters are listed in Table \ref{tab1}.

\begin{table}[]
    \centering
    \caption{Parameters for modeling of the synthetic pre-training dataset.}
    \begin{tabular}{ccc}
         \toprule
         Parameter &  Description & Value \\
         \midrule
         nx & Number of samples in the X axis & 324 \\
         nz & Number of samples in the Z axis & 376 \\
         dx & Sampling step in the X axis & 25 m \\
         dz & Sampling step in the Z axis & 25 m\\
         dt & Recording sampling step & 376 \\
         nt & Number of recording timesteps & 1.6e-2 s \\
         T & Total recording time (nt $\times$ dt) & 6.016 s \\
         nr & Number of receivers & 324 \\
         ds & Receiver spacing  & 25 m \\
         \bottomrule
    \end{tabular}
    \label{tab1}
\end{table}

In the pre-training phase, we generate a total of 2456 training samples for each task. As our framework simultaneously trains on four SPTs (denoising, backscattered noise attenuation, interpolation, and low-frequency extrapolation), the overall dataset comprise 9824 training samples. We employ the AdamW optimizer with a fixed learning rate of $1e-4$ and a batch size of 5. To enhance the stability of the diffusion model training process, we apply an exponential moving average (EMA) with a rate of 0.999. The pre-training is conducted over 200,000 iterations. 

For a fair comparison, the two benchmark models use the same training configuration, except for the EMA mechanism, as their training processes are stable and do not require it. Once pre-training is completed, we evaluate the performance of the pre-trained GSFM and benchmarks on synthetic test data. During inference, to ensure consistency and fairness across all models, we use a single sampling step (i.e., predicting directly at the final step) for generating the synthetic test results. This setup allow us to comprehensively compare the performance of our pre-trained GSFM with the benchmarks. 

\subsection{Denoising}
\begin{figure*}[htbp]
\centering
\includegraphics[width=\textwidth]{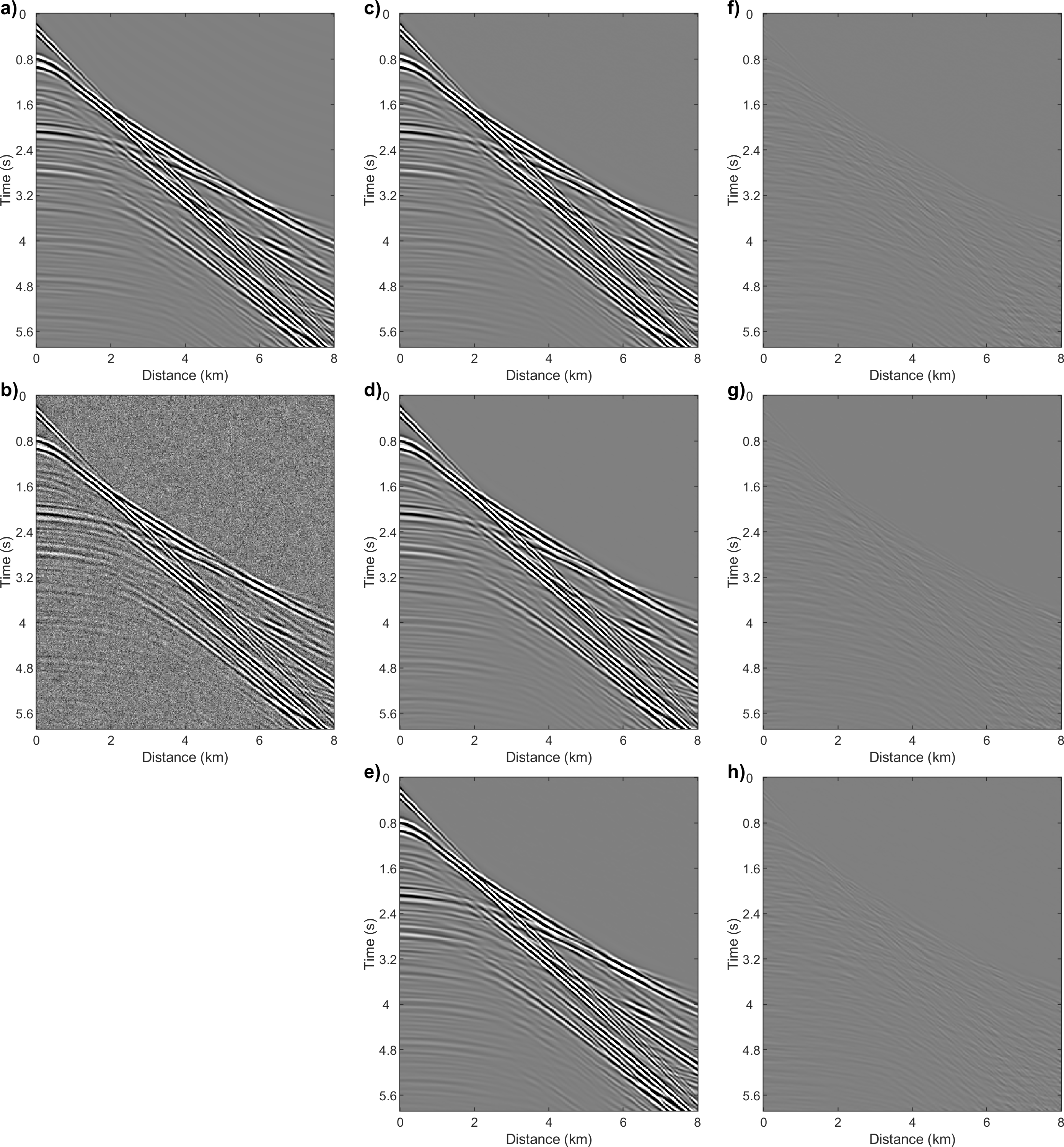}
\caption{Denoising performance comparison between our pre-trained DSFM and two benchmarks on synthetic data. (a) The clean and (b) noisy data, where the noisy data is created by injecting random noise with a level of 30\% into the clean data. The denoised products from (c) our GSFM, (d) Benchmark 1, and (e) Benchmark 2. f, g, and h are the corresponding difference between the denoised results and the clean data. }
\label{fig2}
\end{figure*}

We first test the denoising performance of the pre-trained GSFM on synthetic data contaminated by random noise. Figure \ref{fig2} illustrates the denoising products of the three methods. Panel (a) shows the clean data, while panel (b) represents the noisy test data, which is generated by injecting Gaussian noise with a noise level of 30\%, as follows:
\begin{equation}\label{eq14}
y=x+\epsilon \cdot std(x) \cdot rand(0,1),
\end{equation}
where $\epsilon$ is the noise level, $std(x)$ represents the standard deviation of the clean data $x$, and $rand(0,1)$ is the standard normal distribution. The denoised results for GSFM, Benchmark 1, and Benchmark 2 are displayed in panels (c), (d), and (e) (Figure 2), respectively. The difference between the denoised results and the clean data are presented in panels (f), (g), and (h), respectively. 

Visually, the denoised results from our GSFM and two benchmarks appear very similar, with each method successfully suppressing the random noise and preserving the main seismic reflection events. The differences between the methods are subtle and difficult to evaluate qualitatively, as all methods produce results with comparable reflection continuity and noise suppression. 

To provide a more objective assessment of their performance, Table \ref{tab2} shows a quantitative evaluation of the denoising performance in terms of the MSE metric across different noise levels (10\% to 60\%). The results reveal that GSFM consistently outperforms Benchmark 1 across all noise levels, demonstrating lower MSE values. It is worth noting that Benchmark 1 and GSFM share almost identical architectures, with the only difference being that Benchmark 1 excludes the time encoding module used in the diffusion process. Despite this, GSFM consistently outperforms Benchmark 1 in terms of MSE. This performance gap highlights the significant contribution of GDMs, which contributes to the enhanced denoising performance of the networks. 

Benchmark 2 shows slightly better performance than GSFM at intermediate noise levels (20\% to 50\%), which is likely attributed to one more round of task-specific fine-tuning on labeled data. However, the performance gap is marginal, and GSFM demonstrates superior robustness at the highest noise level. This implies that even without fine-tuning, our pre-trained GSFM has achieved denoising capabilities that match those of the fine-tuned network.

\begin{table}[htbp]
    \centering
    \caption{MSE comparison of denoising performance at different noise levels}
    \begin{tabular}{cccc}
        \toprule
        Noise level & GSFM & Benchmark 1 & Benchmark 2 \\
        \midrule
        10\% &   \textbf{3.09e-07} & 3.44e-07 & 3.21e-07 \\
        20\% &  9.20e-07 & 9.50e-07 & \textbf{9.02e-07} \\
        30\% &  1.67e-06 & 1.73e-06 & \textbf{1.65e-06} \\
        40\% &  2.61e-06 & 2.68e-06 & \textbf{2.59e-06} \\
        50\% &  3.79e-06 & 3.91e-06 & \textbf{3.77e-06} \\
        60\% &  \textbf{4.60e-06} & 4.81e-06 & 4.61e-06 \\
        \bottomrule
    \end{tabular}
    \label{tab2}
\end{table}

\subsection{Backscattered noise attenuation}
\begin{figure*}[htbp]
\centering
\includegraphics[width=\textwidth]{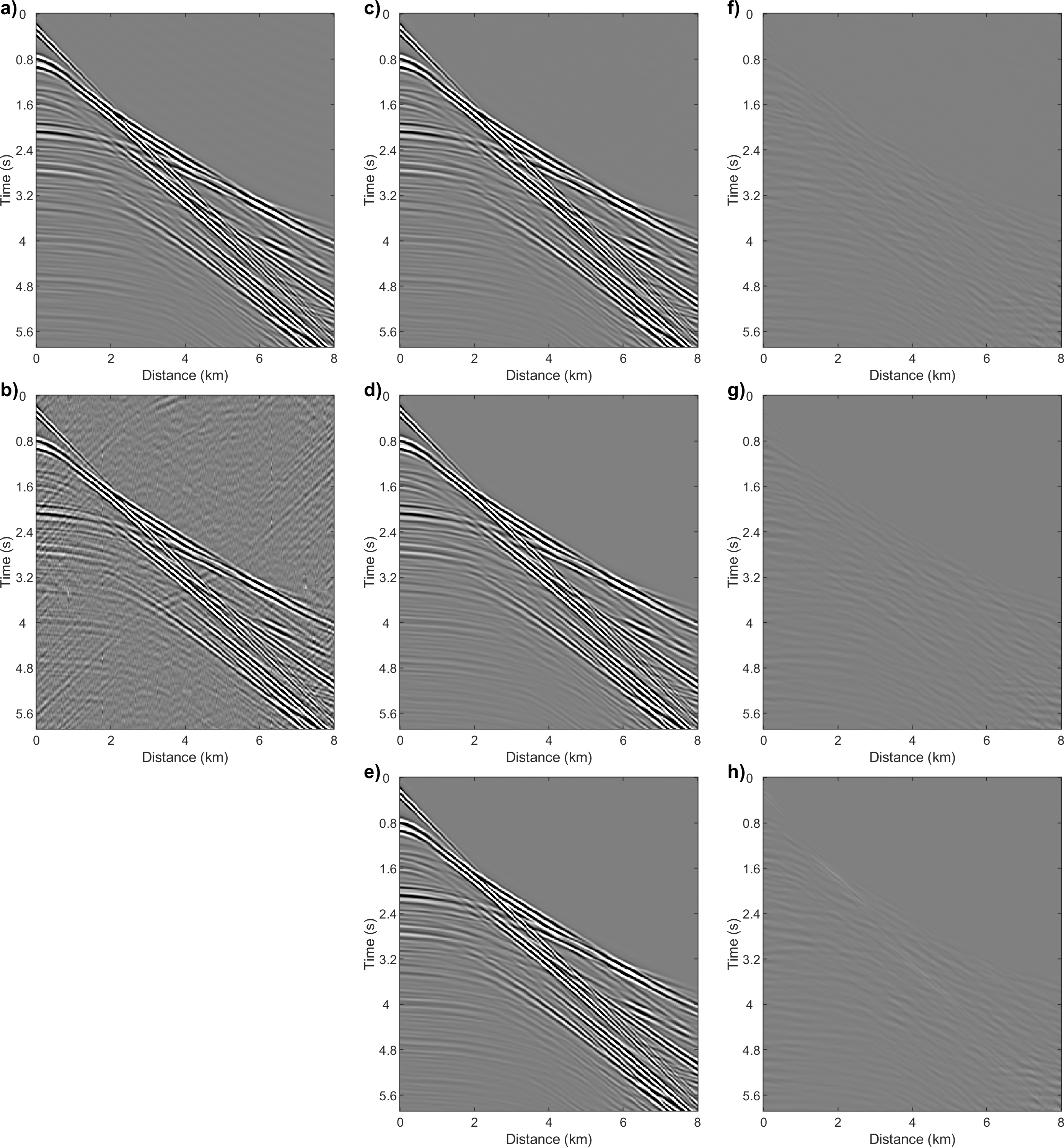}
\caption{Backscattered noise attenuation performance comparison between our pre-trained DSFM and two benchmarks on synthetic data. (a) The clean and (b) noisy data contaminated with backscattered noise. The denoised products from (c) our GSFM, (d) Benchmark 1, and (e) Benchmark 2. f, g, and h are the corresponding difference between the denoised results and the clean data. }
\label{fig3}
\end{figure*}

We, then, evaluate the performance of our pre-trained GSFM in attenuating backscattered noise. Figure \ref{fig3} displays the backscattered noise attenuation results for the three methods. Panel (a) shows the clean seismic data, while panel (b) displays the input data contaminated with backscattered noise. The denoised products from GSFM, Benchmark 1, and Benchmark 2 are presented in panels (c), (d), and (e), respectively, and the corresponding residuals are shown in panels (f), (g), and (h), respectively. 

Similar to the denoising case, the visual differences among the results produced by the three methods are minimal. All methods successfully suppress the backscattered noise and preserve the primary seismic reflections. The residuals reveal that all methods reduce the noise effectively. We further compute the MSE metric for the predicted results of each method. GSFM achieves the lowest MSE of $9.59e-07$, outperforming Benchmark 1 ($1.10e-06$) and Benchmark 2 ($1.26e-06$). This demonstrates GSFM's superior ability to attenuate backscattered noise while preserving the seismic signal.

\subsection{Interpolation}
\begin{figure*}[htbp]
\centering
\includegraphics[width=\textwidth]{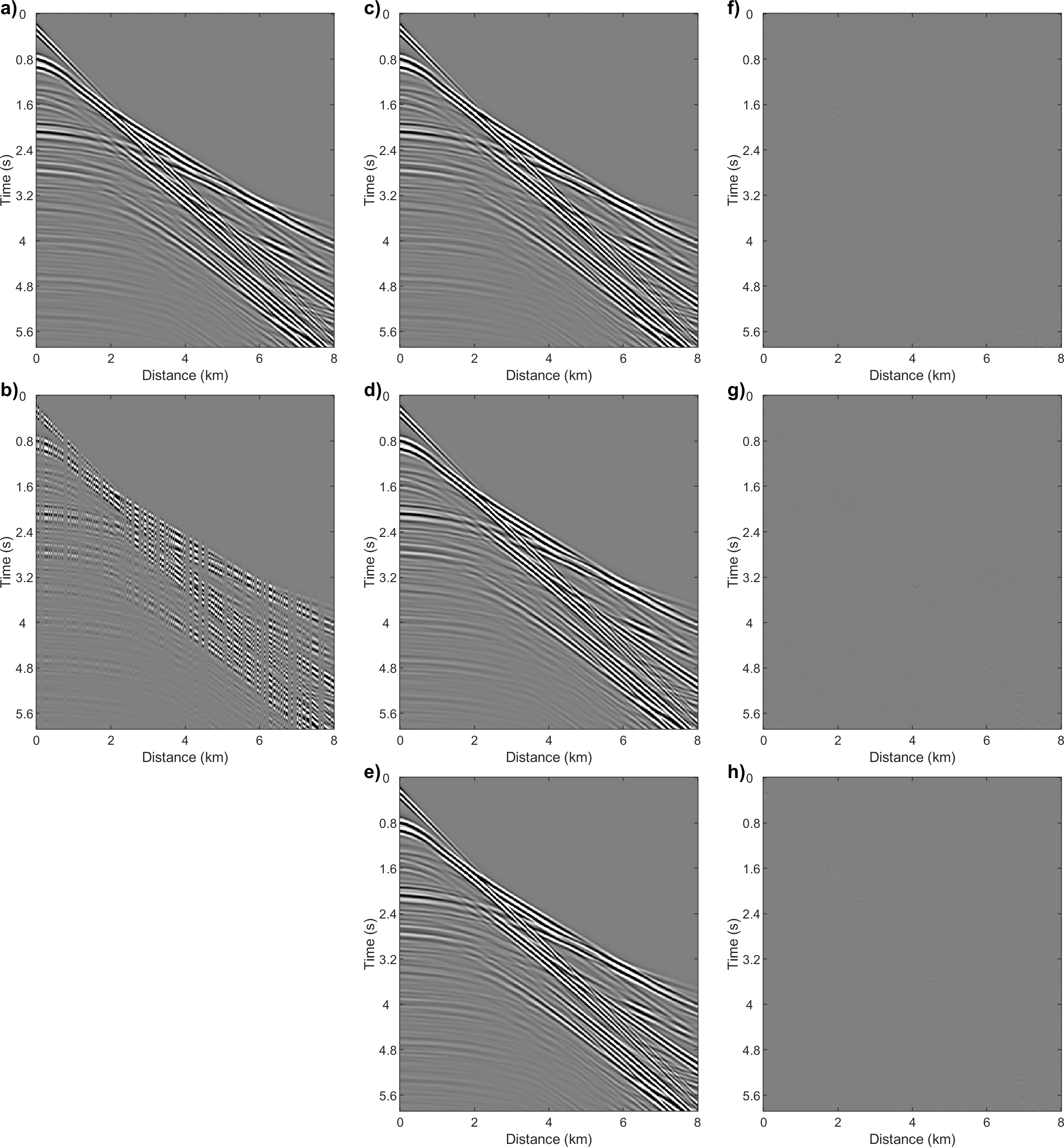}
\caption{Interpolation performance comparison between our pre-trained DSFM and two benchmarks on synthetic data. (a) The complete (label) and (b) incomplete data, where the incomplete data is created by randomly removing 50\% of traces from the complete data. The interpolated products from (c) our GSFM, (d) Benchmark 1, and (e) Benchmark 2. f, g, and h are the corresponding difference between the interpolated results and the labeled data. }
\label{fig4}
\end{figure*}

Furthermore, we evaluate the interpolation performance of our pre-trained GSFM. Figure \ref{fig4} shows the interpolation results for synthetic data with 50\% randomly missing traces. Panel (a) displays the complete labeled data, while panel (b) shows the input data with missing traces. The interpolated results from GSFM, Benchmark 1, and Benchmark 2 are presented in panels (c), (d), and (e), respectively. The corresponding differences between the interpolated results and the complete data are shown in panels (f), (g), and (h), respectively. 

We can see that all three methods achieve visually similar interpolated products, successfully reconstructing the missing traces with less signal leakage. More quantitatively, Table \ref{tab3} summarizes the MSE metrics of the interpolated results across different missing data levels (10\% to 60\%). The results reveal the following trends: 1. At low missing data levels (10\%), GSFM achieves the lowest MSE ($1.45e-08$), outperforming both benchmarks. 2. At intermediate missing data levels (20\% to 50\%), Benchmark 2 slightly outperforms GSFM, demonstrating its effectiveness in handling moderate missing levels due to its task-specific learning on pre-training stage. However, GSFM consistently performs better than Benchmark 1, highlighting its robustness. 3. At the highest missing data level (60\%), GSFM significantly outperforms both benchmarks, achieving an MSE of $3.65e-07$. This result demonstrates GSFM's superior ability to handle highly missing input data. 

Once again, these results demonstrate that the diffusion model boosts the performance of the networks, enabling GSFM to outperform Benchmark 1 consistently. This trend was also observed in the denoising tests, confirming the effectiveness of the diffusion-guided training paradigm. 

\begin{table}[htbp]
    \centering
    \caption{MSE comparison of interpolation performance at different missing levels}
    \begin{tabular}{cccc}
        \toprule
        Missing level & GSFM & Benchmark 1 & Benchmark 2 \\
        \midrule
        10\% &   \textbf{1.45e-08} & 3.60e-08 & 1.73e-08 \\
        20\% &  1.93e-08 & 4.02e-08 & \textbf{1.88e-08} \\
        30\% &  2.85e-08 & 4.51e-08 & \textbf{2.30e-08} \\
        40\% &  8.54e-08 & 8.69e-08 & \textbf{5.21e-08} \\
        50\% &  4.08e-08 & 6.39e-08 & \textbf{3.45e-08} \\
        60\% &  \textbf{3.65e-07} & 5.16e-07 & 5.53e-07 \\
        \bottomrule
    \end{tabular}
    \label{tab3}
\end{table}

\subsection{Low-frequency extrapolation}
\begin{figure*}[htbp]
\centering
\includegraphics[width=\textwidth]{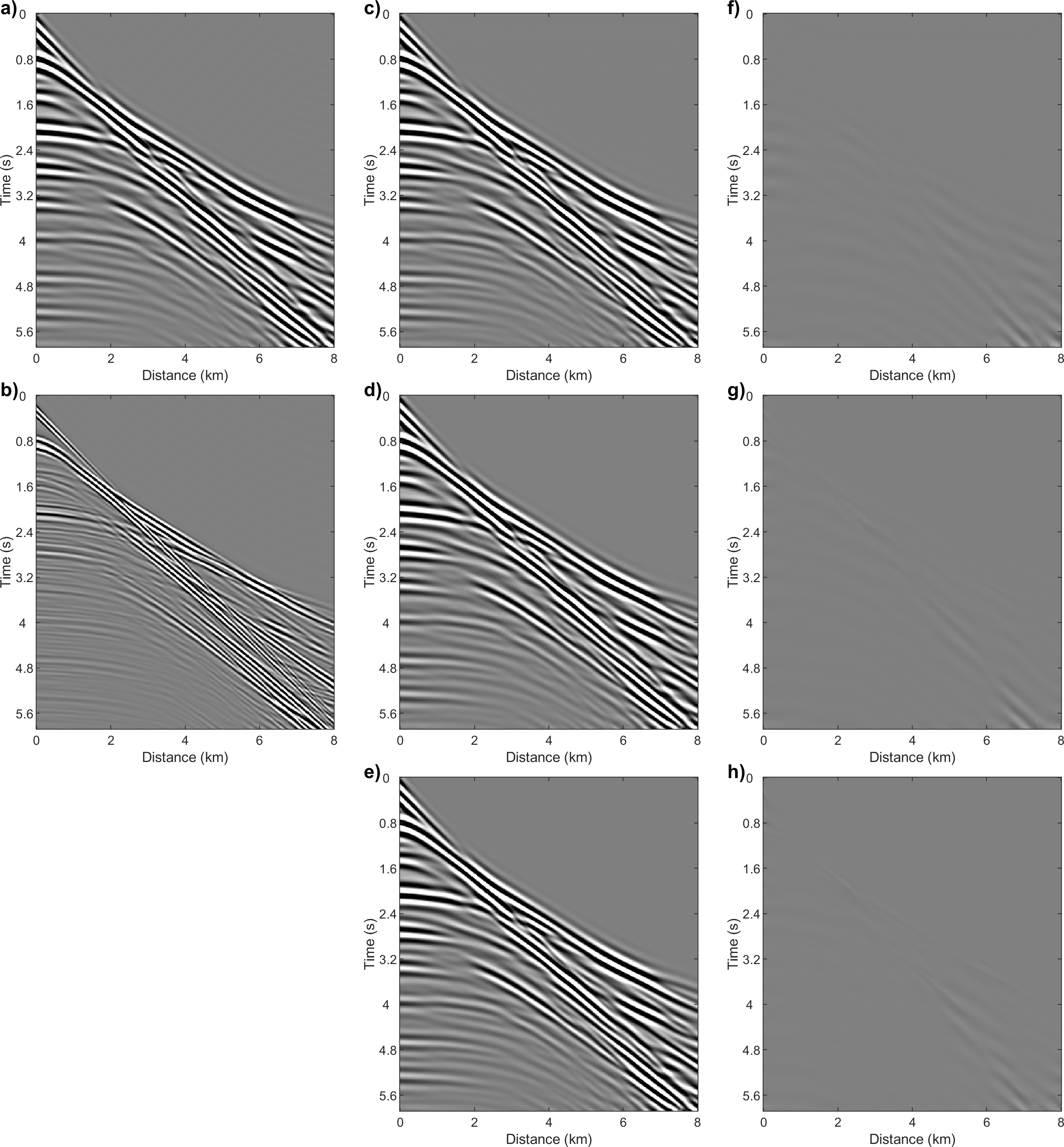}
\caption{Low-frequency extrapolation performance comparison between our pre-trained DSFM and two benchmarks on synthetic data. (a) The labeled and (b) input data, where the input data lacks low frequencies below 4 Hz. The extrapolated products from (c) our GSFM, (d) Benchmark 1, and (e) Benchmark 2. f, g, and h are the corresponding difference between the extrapolated results and the labeled data. }
\label{fig5}
\end{figure*}

Finally, we focus on assessing the capability of our pre-trained GSFM in low-frequency extrapolation, a critical SPT which is particularly beneficial for full-waveform inversion. The extrapolation results for a test data, which miss low-frequencies below 4 Hz, are illustrated in Figure \ref{fig5}. The reference data, including low frequencies, is shown in panel (a), while panel (b) displays the input without low-frequency information. The extrapolated outputs generated by our pre-trained GSFM, Benchmark 1, and Benchmark 2 are depicted in panels (c), (d), and (e), respectively. Panels (f), (g), and (h) highlight the differences between the extrapolated outputs and the reference data. 

Unlike the previous tasks, the difference figures here clearly showcase the differences among the three methods. We can observe that both our GSFM and Benchmark 2 achieve superior extrapolation quality, with minimal residuals and negligible signal leakage. In contrast, Benchmark 1 exhibits more significant signal leakage, particularly near the lower-right region and close to the near-offsets. To complement the visual analysis, we compute the MSE metric between the extrapolated outputs and the reference data for all three methods. GSFM achieve an MSE of $6.0e-07$, while Benchmark 2 slightly outperforms GSFM with an MSE of $3.11e-07$. Benchmark 1, however, performed significantly worse, with an MSE of $1.80e-03$.

\begin{figure*}[htbp]
\centering
\includegraphics[width=\textwidth]{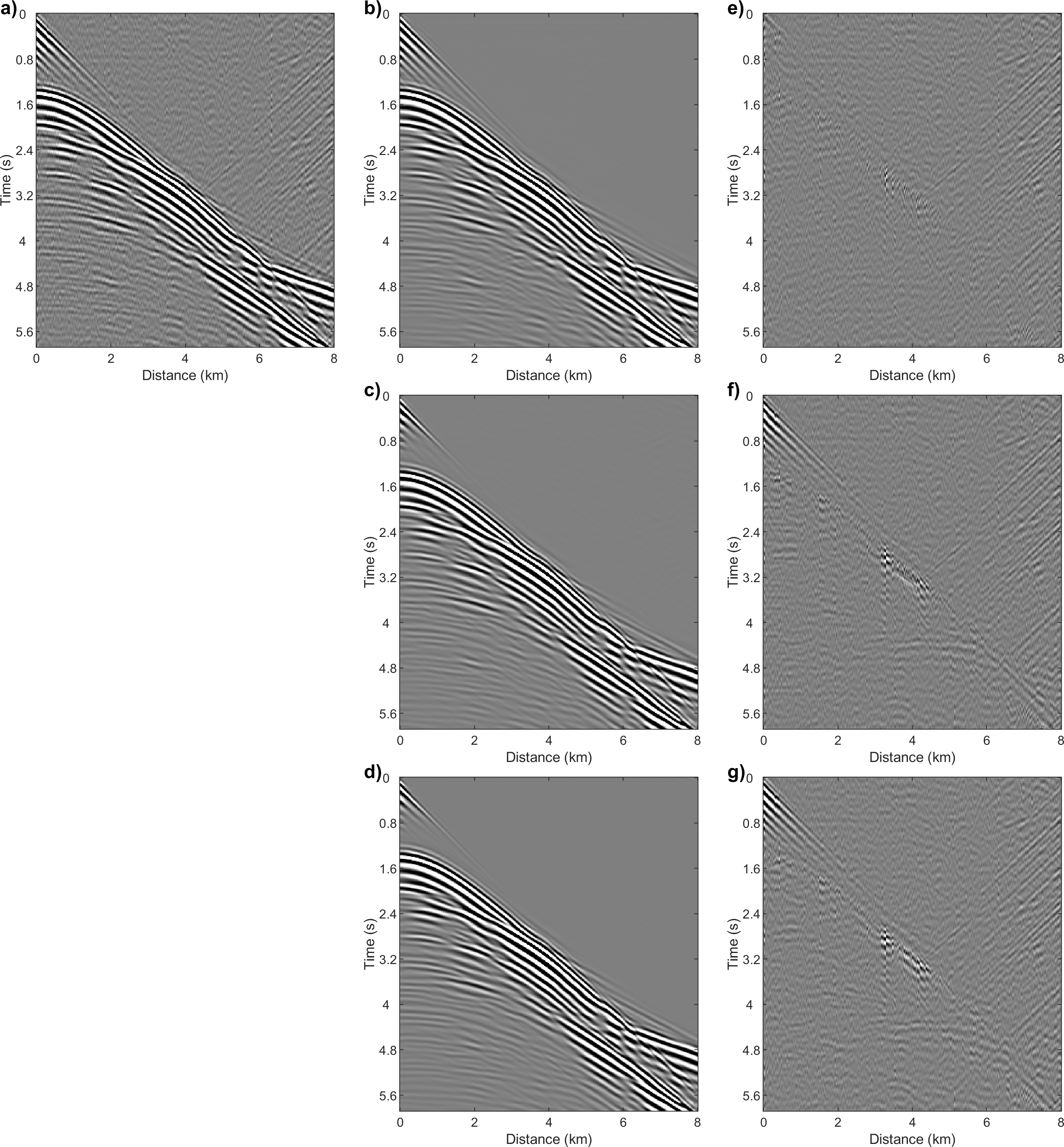}
\caption{Backscattered noise attenuation performance comparison between our fine-tuned DSFM and two benchmarks on field data. (a) The field noisy data contaminated with backscattered noise. The denoised products from (b) our fine-tuned GSFM, (c) Benchmark 1, and (d) Benchmark 2. e, f, and g are he corresponding difference between the denoised results and the field noisy data.}
\label{fig6}
\end{figure*}

\begin{figure*}[htbp]
\centering
\includegraphics[width=\textwidth]{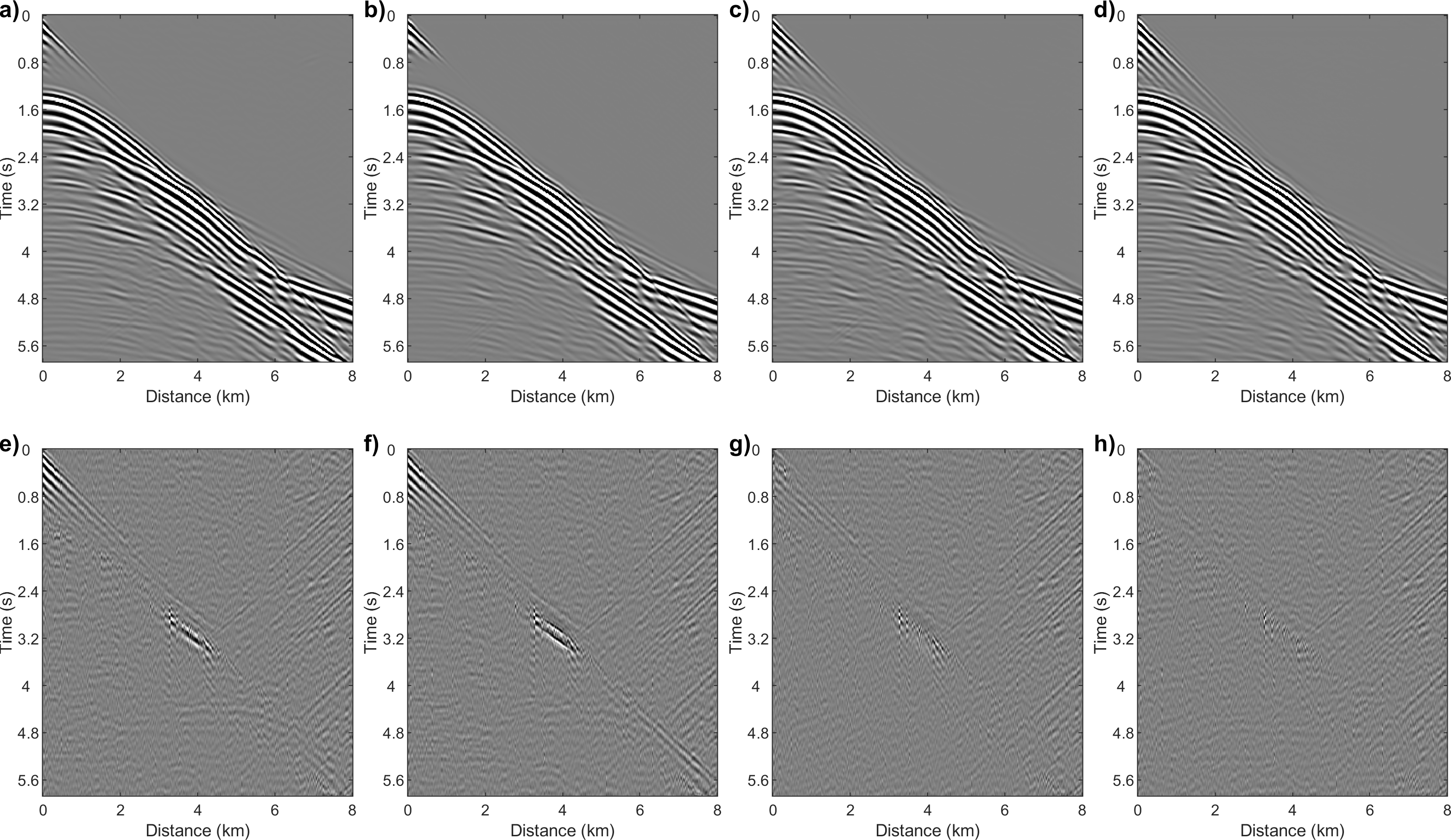}
\caption{Backscattered noise attenuation performance comparison between different fine-tuning strategies. (a) The prediction product from the pre-trained GSFM. The processed products from strategies (b) 1, (c) 2, and (d) 3. e, f, g, and h are the corresponding difference between the processed results and the field noisy data (see Figure \ref{fig6}a). }
\label{fig7}
\end{figure*}

\begin{figure*}[htbp]
\centering
\includegraphics[width=\textwidth]{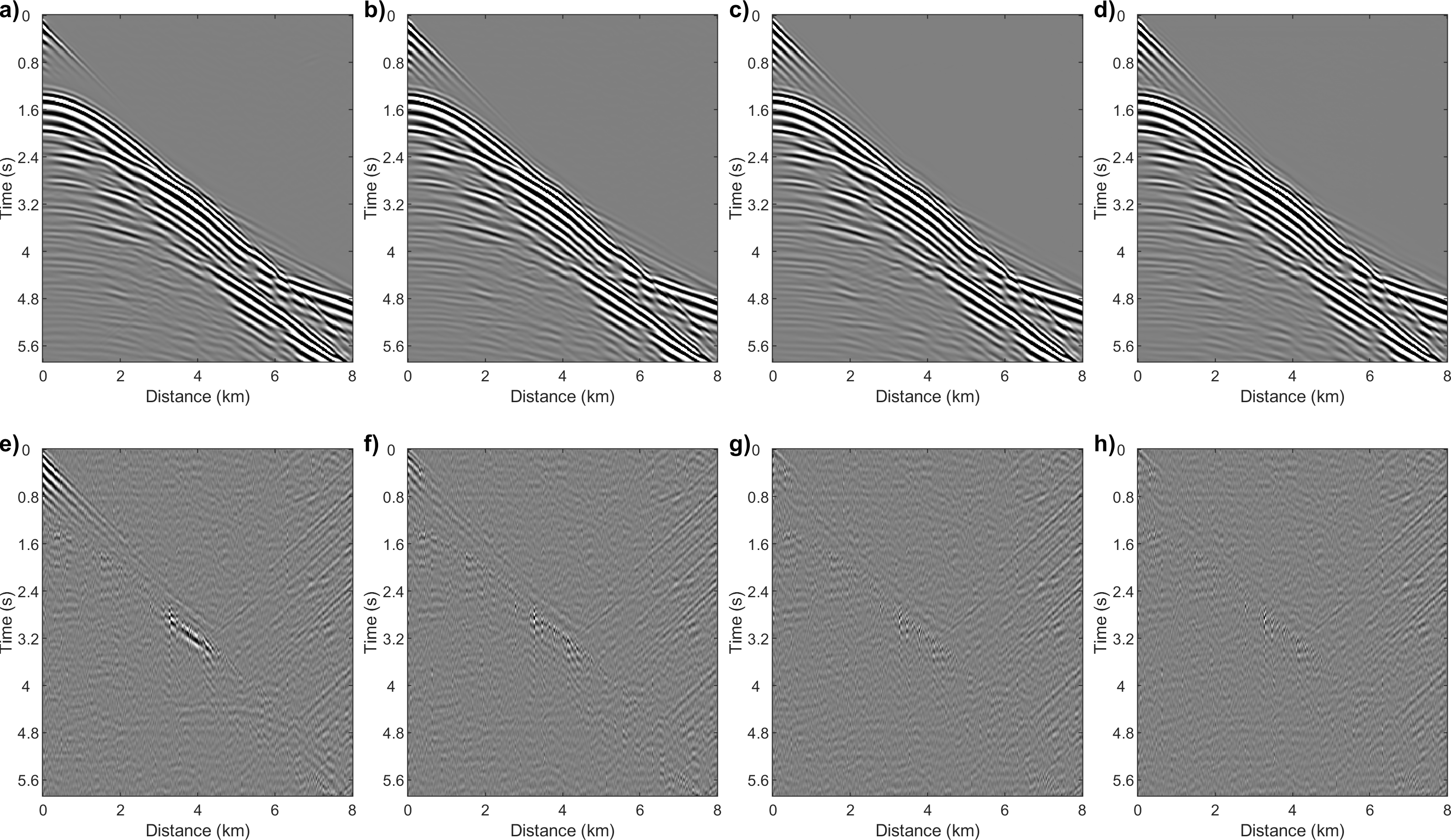}
\caption{Comparison of denoised products of the pre-trained GSFM and the fine-tuned GSFM at different stages. (a) The prediction product from the pre-trained GSFM. (b, c, and d) The prediction products from the fine-tuned stage 1, 5, and 10. e, f, g, and h are the corresponding difference between the processed results and the field noisy data (see Figure \ref{fig6}a). }
\label{fig8}
\end{figure*}

\begin{figure*}[htbp]
\centering
\includegraphics[width=\textwidth]{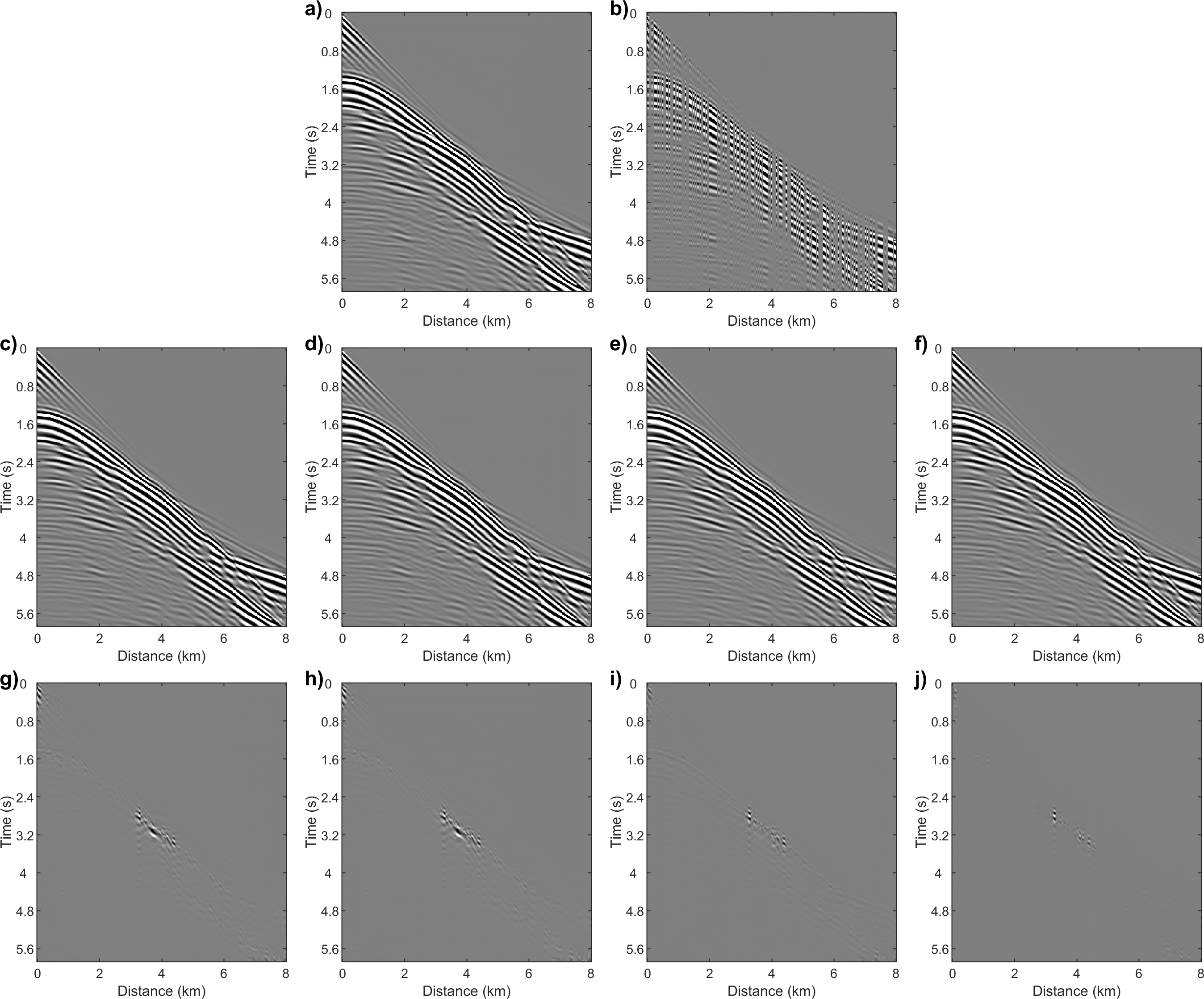}
\caption{Interpolation performance comparison between different fine-tuning strategies. (a) The denoised product from the fine-tuned GSFM on backscattered noise attenuation task, which come from Figure \ref{fig6}b. (b) The incomplete data, where we artificially remove 50\% of the seismic traces from the denoised product. The interpolated products from (c) the pre-trained GSFM and the fine-tuned GSFM using strategies (d) 1, (e) 2, and (f) 3, respectively. g, h, i, and j are the corresponding differences between the interpolated results and the denoised data.}
\label{fig9}
\end{figure*}

\begin{figure*}[htbp]
\centering
\includegraphics[width=\textwidth]{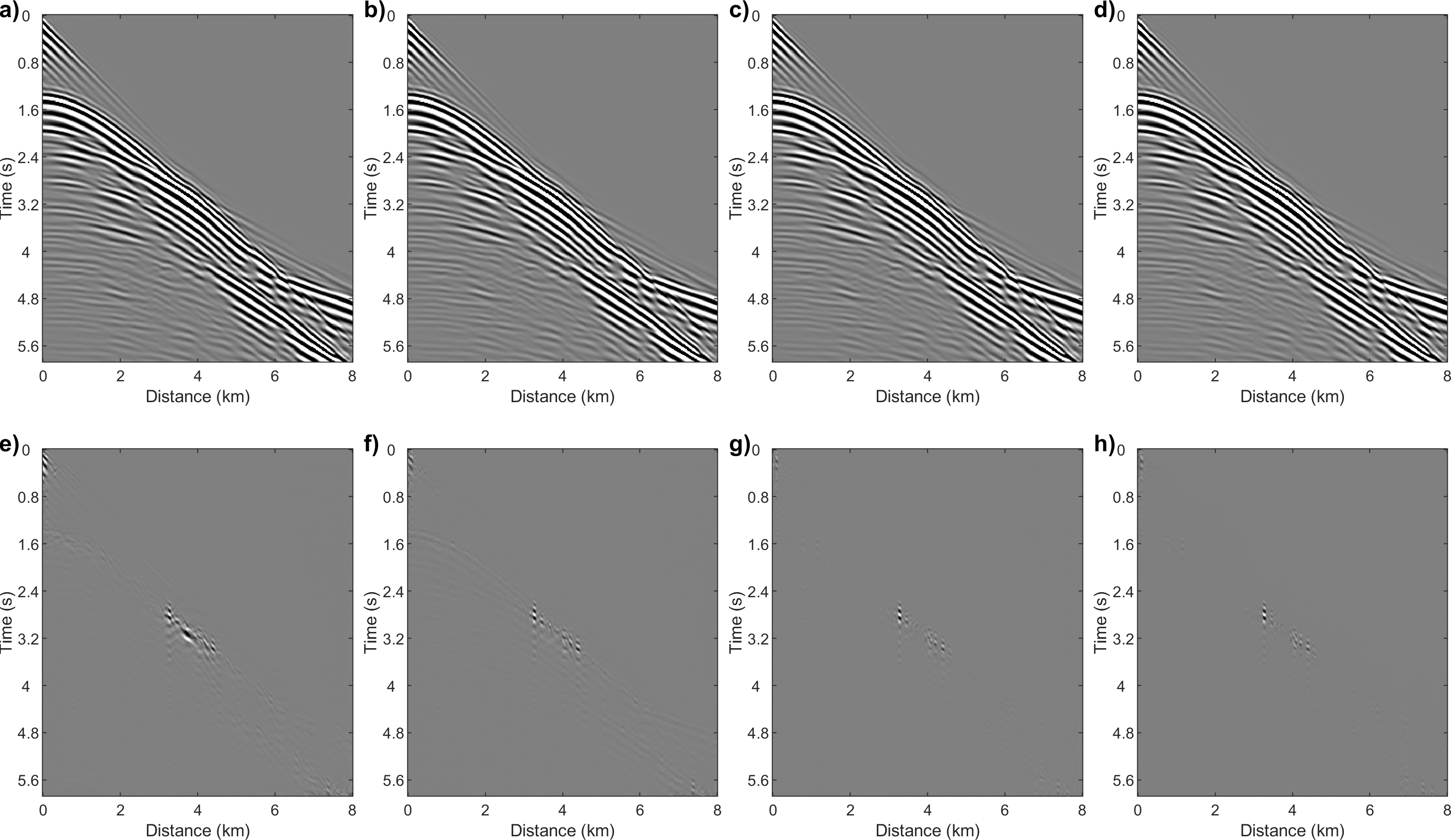}
\caption{Comparison of interpolated products of the pre-trained GSFM and the fine-tuned GSFM at different stages. (a) The interpolated product from the pre-trained GSFM. (b, c, and d) The interpolated products from the fine-tuned GSFM at the stages 1, 5, and 10. e, f, g, and h are the corresponding difference between the interpolated results and the denoised data (see Figure \ref{fig9}a). }
\label{fig10}
\end{figure*}

\subsection{Understanding performance differences among the methods}
The evaluations of our pre-trained GSFM, Benchmark 1, and Benchmark 2 across the four SPTs on synthetic data offer insights into their strengths, limitations, and fundamental differences. 

Benchmark 1, based on the conventional NN paradigm, consistently underperforms compared to GSFM and Benchmark 2. While its performance gap is less pronounced in the first three tasks, it becomes significantly evident in the low-frequency extrapolation task. We know that Benchmark 1 approximates the nonlinear relationship between the input and target data. For the first three tasks, the target data (clean, complete seismic data) remains consistent across tasks. This consistency enables the network to learn a more generalized mapping that performs adequately across these tasks. However, in the low-frequency extrapolation task, the target data shifts to clean, full-band seismic data, including low frequencies that are absent in the input. This change introduces a more specific and challenging relationship to learn, which the traditional paradigm struggles to approximate effectively. As a result, the network is biased towards learning a more generalized mapping, leading to insufficient focus on the specific relationship required for accurate low-frequency extrapolation. 

In contrast, GSFM, despite sharing the similar architecture as Benchmark 1, leverages the GDMs to capture and learn a more unified distribution. By modeling the joint distribution of clean, complete, and full-band seismic data, GSFM is able to bridge the gap between the input and target data more effectively, enabling it to achieve more accurate and robust results across a broader range of tasks. 

Benchmark 2 consistently demonstrates strong performance across tasks even slightly outperforms GSFM in terms of MSE for certain tasks. However, this slight advantage is achieved through task-specific fine-tuning, which relies heavily on labeled datasets and requires additional computational resources. Although additional fine-tuning can improve performance on synthetic data, our ultimate goal is field data. Since Benchmark 2 conducts fine-tuning using synthetic data, it still faces generalization challenges. In contrast, our GSFM does not rely on task-specific fine-tuning with labeled data. Instead, it undergoes fine-tuning directly on field data in an SSL manner, enabling it to address the generalization challenges faced by Benchmark 2. In the following section, we will share our field data test to highlight these advantages, showcasing GSFM’s effectiveness in addressing the complexities of diverse SPTs. 

\section{\textbf{Field data examples}}
In this section, we will go forward to fine-tune the pre-trained GSFM on real data and, then, evaluate the performance of our fine-tuned GSFM on field data across denoising, interpolation, and low-frequency extrapolation tasks. Also, we will examine the effectiveness of three different fine-tuning strategies, as outlined in the Fine-tuning subsection of the Method section. Finally, we discuss how GSFM can be leveraged for uncertainty quantification in seismic processing and, also, illustrate how use uncertainty quantification to guide our fine-tuning process.

\subsection{Field data and Fine-tuning configuration}
We use a marine field dataset to test our method. This dataset was acquired using a steamer survey in North West Australia. The original dataset consists of 1824 shot gathers activated with air gun sources, with an approximate horizontal spacing of 18.75 m and a sampling rate of 1 ms. Each shot gather contains 648 receivers, spaced 12.5 m apart. For testing purposes, we select every third shot gather starting from the left, resulting in a total of 200 shot gathers. To reduce the computational burden during training, the number of receivers in the field data was reduced to 324, and the time samples are downsampled from 6016 to 376, following the preprocessing used in \cite{harsuko2024optimizing}. 

During fine-tuning on field data using the three different strategies, we ensure a fair comparison by using the same total number of iterations, set to 30000. However, in Strategy 3, these 30000 iterations are divided into 10 stages. Specifically, in Algorithm \ref{alg1}, the total stages $S$ are set to 10, and the iterations per stage $N_{stage}$ are set to 3000. We emphasize that all additional fine-tuning configurations remain consistent. Specifically, the learning rate is fixed at $5e-5$, the batch size is set to 4, and the EMA rate is configured to 0.999. Furthermore, during the sampling process, the pseudo-label generation utilized a diffusion step respace of 1 from the original 1000 diffusion steps. This means that only a single sampling step is used for field data predictions, which significantly improves fine-tuning and processing efficiency to meet the demands of practical applications. 

We also emphasize that, since the field data does not include random noise, we do not fine-tune the pre-trained GSFM for the denoising task in this case. Instead, we independently optimize the pre-trained GSFM for the backscattered noise attenuation, interpolation, and low-frequency extrapolation tasks. During fine-tuning, we adopt a sequential workflow similar to the traditional seismic processing paradigm. Specifically:
\begin{itemize}
    \item \textbf{1. Backscattered noise attenuation}: Fine-tuning is first applied to this task to address inherent noise in the field data. Here, the noise added to pseudo labels is extracted from the area outside the first arrival.
    \item \textbf{2. Interpolation}: After obtaining denoised results, since the denoised data does not contain bad traces, we artificially remove 50\% of the seismic traces from the data to construct the incomplete seismic data, serving as the original fine-tuning dataset for interpolation task. To better reflect practical scenarios, where certain fixed receivers in a steamer are damaged, the indices of the missing 50\% traces hold same across the selected 200 shot gathers.
    \item \textbf{3. Low-Frequency extrapolation}: Finally, the denoised data is used as our initial training data for fine-tuning the GSFM to perform low-frequency extrapolation.
\end{itemize}

This sequential workflow ensures that each task builds upon the results of the previous task, aligning with real-world seismic processing practices. By fine-tuning GSFM directly on field data in an SSL manner, we aim to demonstrate its capability to address generalization challenges, particularly when labeled data is unavailable. In the subsequent subsections, we present the results for each task. 

\begin{figure*}[htbp]
\centering
\includegraphics[width=0.75\textwidth]{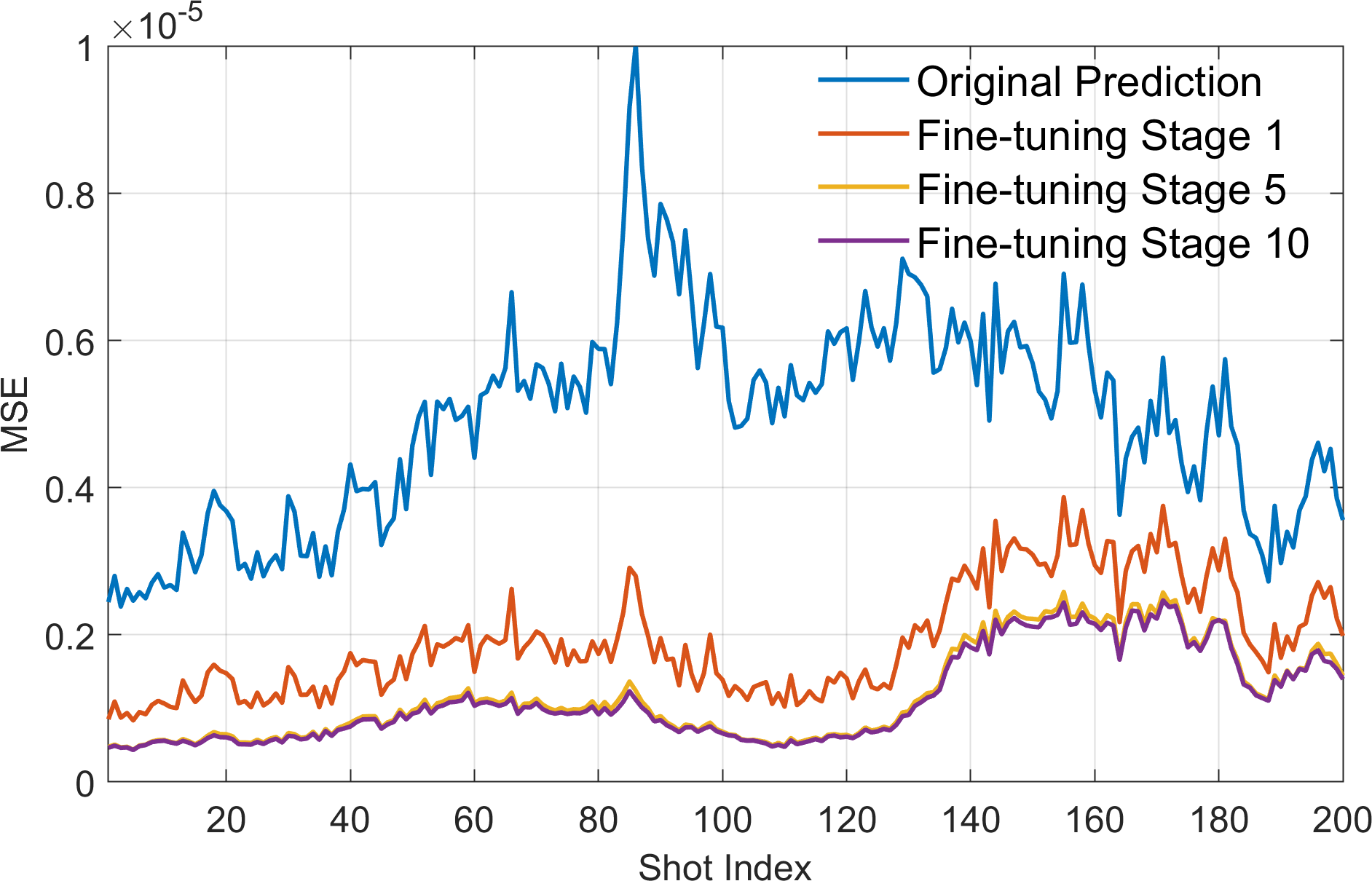}
\caption{The MSE metric of the interpolation results of the selected 200 incomplete shot gathers from the pre-trained GSFM and the fine-tuned GSFM at different stages. The 200 incomplete shot gathers are obtained by removing a fixed 50\% of the traces from the denoising shot gathers. The Original Prediction legend represents the prediction results of the pre-trained GSFM for 200 incomplete shot gathers. The Fine-tuning stage 1, 5, and 10 legends correspond to the prediction results of the GSFM fine-tuned at stages 1, 5, and 10 for 200 incomplete shot gathers, respectively.}
\label{fig11}
\end{figure*}

\begin{figure*}[htbp]
\centering
\includegraphics[width=0.7\textwidth]{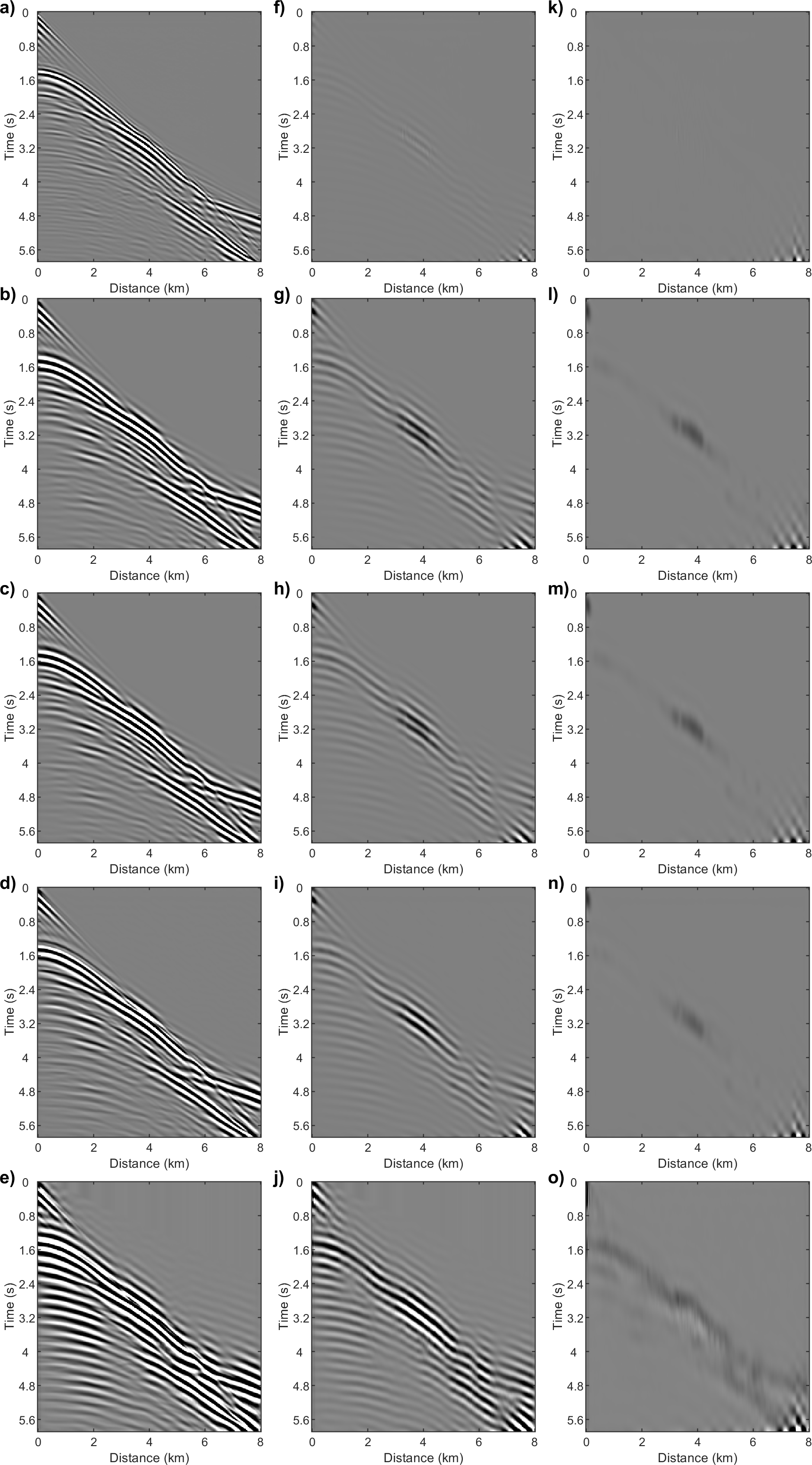}
\caption{Low-frequency extrapolation performance comparison between different fine-tuning strategies. (a) The denoised product from the fine-tuned GSFM on backscattered noise attenuation task, which come from Figure \ref{fig6}b. The predicted products using (b) pre-trained GSFM and the fine-tuned GSFM using strategies (c) 1, (d) 2, and (e) 3, respectively. The second and third columns correspond to frequency components less than 4 Hz and 2 Hz, respectively. }
\label{fig12}
\end{figure*}

\begin{figure*}[htbp]
\centering
\includegraphics[width=0.8\textwidth]{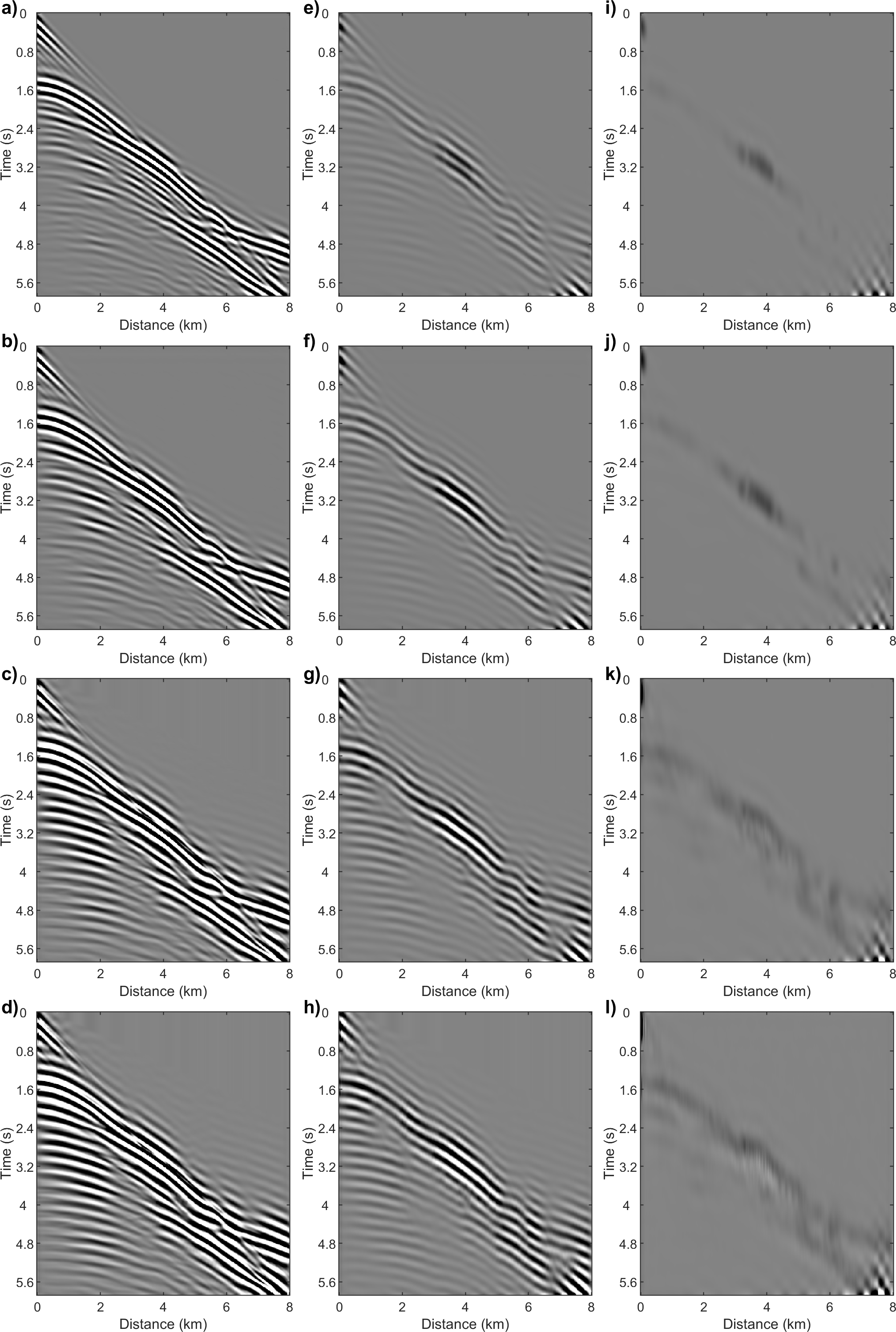}
\caption{Comparison of low-frequency extrapolation products of the pre-trained GSFM and the fine-tuned GSFM at different stages. (a) The extrapolated product from the pre-trained GSFM. b, c, and d are the extrapolated products from the fine-tuned GSFM at the stages 1, 5, and 10. The second and third columns correspond to frequency components less than 4 Hz and 2 Hz, respectively.}
\label{fig13}
\end{figure*}

\subsection{Backscattered noise attenuation}
we first evaluate the performance of our fine-tuned GSFM for backscattered noise attenuation on field data. We compare the outputs of our method against those from Benchmark 1 and Benchmark 2, which is detailed in last section. Furthermore, we explore the impact of various fine-tuning strategies and highlight the iterative improvements achieved with Strategy 3. 

Figure \ref{fig6} illustrates the comparison of the denoised results between our fine-tuned GSFM and the two benchmarks. Panel (a) shows the field data contaminated with backscattered noise, while panels (b), (c), and (d) present the denoised outputs produced by our fine-tuned GSFM, Benchmark 1, and Benchmark 2, respectively. The corresponding differences between the denoised results and the noisy field data are shown in panels (e), (f), and (g), respectively. It is evident that both benchmarks fail to suppress the backscattered noise effectively, leading to severe signal leakage in their outputs. Specifically, compared with Benchmarks 1, Benchmarks 2 exhibit more significant signal leakage. In contrast, our fine-tuned GSFM demonstrates superior performance, successfully preserving the true signal while significantly reducing the noise. This highlights GSFM’s capability to adapt to field data through SSL fine-tuning, addressing the generalization challenges faced by the benchmarks. 

To evaluate the influence of different fine-tuning strategies, we compare the processed outputs in Figure \ref{fig7}. Panel (a) illustrates the result from the pre-trained GSFM before any fine-tuning, while panels (b), (c), and (d) show the outputs after fine-tuning with strategies 1, 2, and 3, respectively. The corresponding differences are shown in panels (e), (f), (g), and (h). The results reveal distinct differences between the strategies. Strategy 1 shows minimal improvement over the pre-trained GSFM, however in some areas, it even increases signal leakage. Strategy 2 provides moderate enhancements in noise reduction, yet residuals persist. Strategy 3, however, achieves substantial gains, effectively reducing noise and preserving the signal structure. This demonstrates the importance of a progressive fine-tuning process in adapting the model to complex field data. 

To further better understand the iterative improvement offered by Strategy 3, Figure \ref{fig8} presents the outputs at different fine-tuning stages. Panel (a) displays the pre-trained GSFM's output, while panels (b), (c), and (d) show the results from stages 1, 5, and 10 of fine-tuning. The corresponding differences are displayed in panels (e), (f), (g), and (h). The stepwise refinement process is clearly evident in these results. Early in the fine-tuning phase, such as stage 1, signal leakage remain prominent. By stage 5, the model demonstrates reduced signal leakage. At the final stage, stage 10, the processed output is of high quality, with excellent noise attenuation and signal preservation. This progressive refinement process reflects the iterative fine-tuning strategy's core strength: it allows GSFM to gradually capture the distribution of the field data, shifting its understanding from the synthetic data distribution toward the field data distribution. This adaptation is critical for real-world applications, as it ensures that the model generalizes effectively to field data even without labeled ground truth. 

\subsection{Interpolation}
We, then, explore the interpolation capabilities of our pre-trained GSFM and its fine-tuned versions when applied to field data. Different fine-tuning strategies are assessed to identify the most effective approach. Moreover, we analyze how the iterative refinement process enhances interpolation performance at successive stages. 

The interpolation results obtained using the pre-trained GSFM and the fine-tuned GSFM under different strategies are depicted in Figure \ref{fig9}. Panel (a) presents the denoised product from the backscattered noise attenuation task (Figure \ref{fig6}b), while panel (b) displays the incomplete data created by removing 50\% of the seismic traces. Panels (c), (d), (e), and (f) show the outputs of the pre-trained GSFM and the models fine-tuned using strategies 1, 2, and 3, respectively. The differences for each result relative to the original denoised data are illustrated in panels (g), (h), (i), and (j). The results show noticeable variations in performance across the strategies. Strategy 1, which minimally adjusts the pre-trained model, produces results that closely resemble the pre-trained GSFM’s output, leaving significant interpolation gaps unaddressed. Strategy 2 improves interpolation performance by reducing gaps, yet some residual inaccuracies remain. Strategy 3, in contrast, delivers the most refined results, reconstructing the missing traces with superior accuracy and significantly reducing errors. This progression highlights the advantage of iterative fine-tuning for aligning the model with field data. 

To better understand the iterative refinement enabled by Strategy 3, Figure \ref{fig10} displays the interpolated results at different stages of fine-tuning. The initial result from the pre-trained GSFM is presented in panel (a), while panels (b), (c), and (d) show the outputs after stages 1, 5, and 10 of fine-tuning, respectively. The corresponding differences are displayed in panels (e), (f), (g), and (h). With each fine-tuning stage, the quality of interpolation improves. At stage 1, the model addresses some missing traces but leaves substantial residuals. By stage 5, the interpolation becomes more accurate, with reduced signal leakage. Stage 10 achieves optimal results, effectively reconstructing the missing traces with minimal discrepancies. This progression demonstrates the power of Strategy 3 in leveraging iterative updates to adapt the model's predictions to the field data distribution. 

Since the incomplete field data is generated by artificially removing 50\% of the seismic traces from the denoised data, we can obtain corresponding labeled data. As a result, we can provide a quantitative perspective to evaluate the interpolation performance across all 200 shot gathers at different fine-tuning stages. The mean squared error (MSE) for interpolation is plotted in Figure \ref{fig11}. The curve labeled "Original Prediction" represents the pre-trained GSFM, while the curves for stages 1, 5, and 10 correspond to the respective fine-tuning iterations. The MSE trends reveal a clear improvement as fine-tuning progresses. Compared to the pre-trained GSFM, stage 1 achieves a noticeable reduction in MSE. By stage 5, the interpolation accuracy improves further, with MSE values continuing to decline. Stage 10 marks the culmination of the process, yielding the lowest MSE across all shot gathers. This trend underscores the effectiveness of the iterative refinement process in progressively aligning the model with the complexities of field data. 

\subsection{Low-frequency extrapolation}
We, finnaly, assess the performance of our pre-trained and fine-tuned GSFM models on the low-frequency extrapolation task for field data. By analyzing results across different fine-tuning strategies, we demonstrate how the iterative refinement process enables the model to progressively capture missing low-frequency components in real data. 

Figure \ref{fig12} presents a detailed comparison of the low-frequency extrapolation results produced by the pre-trained GSFM and models fine-tuned with Strategies 1, 2, and 3. Panel (a) displays the denoised product obtained from the backscattered noise attenuation task (see Figure \ref{fig6}b), which serves as the input for extrapolation. Panels (b), (c), (d), and (e) show the results from the pre-trained GSFM and the fine-tuned models using Strategies 1, 2, and 3, respectively. To provide a focused view of the low-frequency content, the second and third columns highlight frequency components below 4 Hz and 2 Hz, respectively, where we use the low-pass filter to leave the low-frequency component. The results clearly illustrate the progression in reconstruction quality across the strategies. Strategy 1 offers only negligible improvements over the pre-trained GSFM. Strategy 2 delivers better results, reconstructing slight more of the missing low-frequency details. However, it struggles to effectively recover frequencies below 2 Hz. In contrast, Strategy 3 achieves the most substantial enhancement, recovering nearly complete low-frequency components both below 4 Hz and 2 Hz. This demonstrates that Strategy 3 is effective in adapting the model to field data and recovering challenging low-frequency details. 

Again, to investigate the refinement process under Strategy 3, Figure \ref{fig13} illustrates the extrapolated outputs at different fine-tuning stages. Panel (a) represents the prediction from the pre-trained GSFM, while panels (b), (c), and (d) show the results at fine-tuning stages 1, 5, and 10, respectively. The corresponding frequency components below 4 Hz and 2 Hz are highlighted in the second and third columns. The progression across stages demonstrates the strength of iterative refinement. In the early stages, the model begins to reconstruct low-frequency content, but the frequencies below 2 Hz remain less apparent. By stage 5, improvements become evident. At stage 10, the model delivers its best performance, with nearly complete recovery of low-frequency components, including those below 2 Hz. This gradual enhancement further underscores how Strategy 3 enables the GSFM to iteratively align with the field data distribution and, thus, improve the performance of GSFM on real data. 

\begin{figure*}[htbp]
\centering
\includegraphics[width=\textwidth]{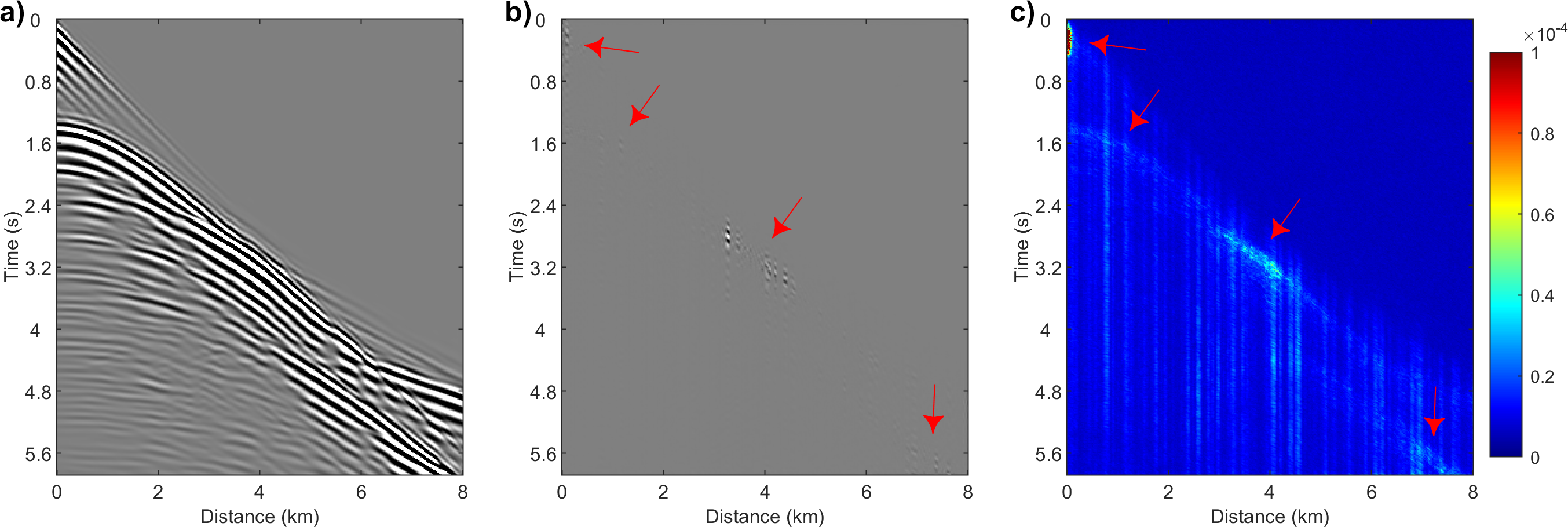}
\caption{The mean of multiple interpolation results under same condition and the corresponding uncertainty, where the condition is the incomplete real data (see Figure \ref{fig9}b). (a) The mean of multiple interpolation results. (b) The differences between the mean of multiple interpolation results and the labeled data, where the labeled data is the denoised data (see Figure \ref{fig6}b). (c) The uncertainty of interpolation result.}
\label{fig14}
\end{figure*}

\begin{figure*}[htbp]
\centering
\includegraphics[width=0.8\textwidth]{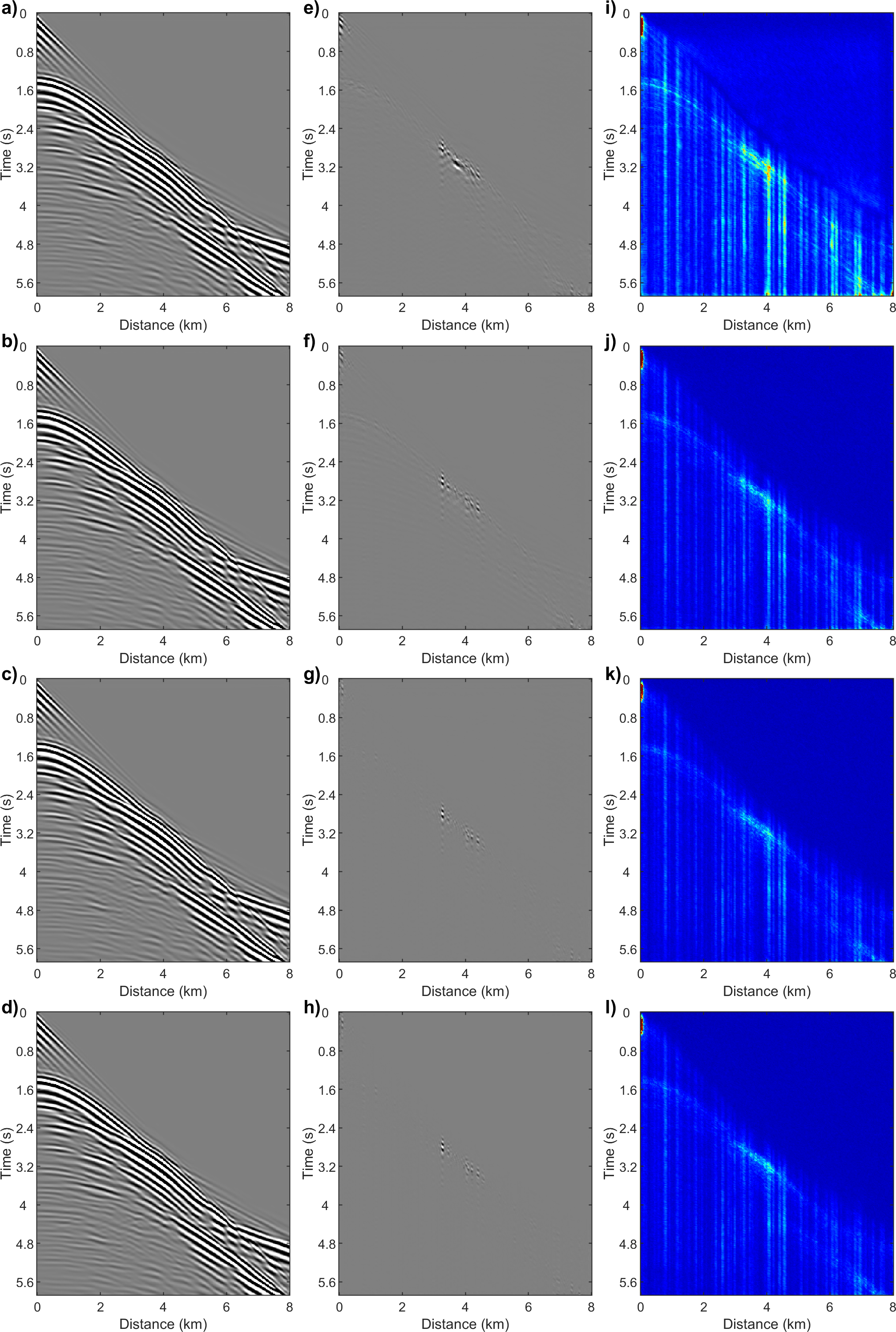}
\caption{The uncertainty dynamics at fine-tuning process for field data interpolation. (a, b, c, d) The mean of 50 interpolation predictions generated by the pre-trained GSFM, and the fine-tuned GSFM at stages 1, 5, and 10, respectively. (e, f, g, h) The difference between the mean interpolation results and the labeled data (denoised results) at the corresponding stages. (i, j, h, l) The uncertainty calculated as the standard deviation of 50 predictions for each model at the corresponding stages.}
\label{fig15}
\end{figure*}

\subsection{Uncertainty quantification}
Traditional NN-based seismic processing paradigms and, also, pre-training and fine-tuning strategies lack the ability to effectively quantify the uncertainty of their processing products. In practical applications, uncertainty quantification is critical as it provides insight into the reliability of the processing results, aiding in decision-making processes. Due to the inherent probabilistic nature of GDMs, they naturally lend us to estimating the uncertainty of processing results. Although our GSFM employs a deterministic sampling process by targeting $x_0$ and setting $\sigma_t=0$, we emphasize that the sampling process still originates from random noise. Therefore, when the seed for sampling is randomized, the initial random noise varies, leading to slight differences in the generated processing results. Consequently, we can still effectively perform uncertainty quantification. 

The method for uncertainty quantification is straightforward. The input condition, e.g., the incomplete seismic data, is replicated $B$ times along the batch dimension. Then, $B$ different random noise are sampled, and the GSFM processes these inputs to generate $B$ corresponding predictions. By calculating the mean of these $B$ predictions along the batch dimension, we obtain the final predicted result. Simultaneously, the standard deviation of the $B$ predictions along the batch dimension provides an indication of variance that can be used for uncertainty quantification. 

To illustrate this, we conduct a test on the interpolation task using field data. Specifically, the incomplete field data (see Figure \ref{fig9}b) is replicated 50 times along the batch dimension, and 50 different random noise samples are used during the prediction process. As a result, we obtain 50 interpolation results, where we use the GSFM fine-tuned at the 10th stage. The mean and standard deviation of these predictions are shown in Figure \ref{fig14}. Panel (a) shows the mean of 50 interpolation results, while panel (b) depicts the difference between the mean interpolation result and the labeled data (see Figure \ref{fig6}b). Panel (c) denotes the uncertainty of interpolation result. We can see that the regions with significant signal leakage exhibit higher uncertainty values, as indicated by the red arrows. This correlation demonstrates that the calculated uncertainty effectively identifies regions in the processed results with higher errors. We emphasis that for the incomplete data used here, derived by removing 50\% of traces from the denoised data, the labeled data is available for direct comparison. This allows us to identify the signal leakage and assess each model’s performance. However, in practice, for field data without available labels, evaluating the prediction quality becomes challenging. In such cases, the uncertainty quantification (Panel (c)) provides a valuable alternative. Such information is invaluable for assessing the quality and reliability of seismic processing results, significantly enhancing the confidence in subsequent applications. 

Actually, the correlation between reduced uncertainty and improved interpolation performance further inspires us to explore whether such uncertainty measures can guide the fine-tuning process. To demonstrate this, Figure \ref{fig15} further explores the uncertainty quantification across pre-trained and GSFM at different fine-tuning stages (stages 1, 5, and 10). Each row in Figure \ref{fig15} corresponds, from top to bottom, to the pre-trained GSFM and the GSFM fine-tuned at the 1st, 5th, and 10th stages, respectively. The columns, from left to right, represent the mean of the predictions from 50 samples, the difference between the mean prediction and the labeled data, and the uncertainty. 

From Figure \ref{fig15}, we can observe the following trends:
\begin{itemize}
    \item \textbf{Prediction consistency}: Across all stages, the mean prediction results appear visually similar, indicating that noticeable performance differences are not easily discernible without a direct comparison of residuals or uncertainty.
    \item \textbf{Reduction in signal leakage}: The residual figures show that as the fine-tuning progresses (from stage 1 to stage 10), the signal leakage consistently decreases, confirming the improvement in interpolation accuracy.
    \item \textbf{Uncertainty dynamics}: The standard deviation measure figures reveal a steady reduction in uncertainty as fine-tuning progresses. This decrease aligns with the reduced signal leakage observed in the residuals, emphasizing the strong relationship between uncertainty and performance.
\end{itemize}

Therefore, these findings highlight the potential of using uncertainty as a guiding metric during fine-tuning stage. Specifically, we can monitor the reduction in uncertainty to evaluate whether fine-tuning is effectively optimizing the model. If the uncertainty stabilizes and shows no significant reduction across successive fine-tuning iterations, it may indicate convergence, suggesting that further fine-tuning is unnecessary.

\section{Discussion}
This section provides an in-depth analysis of various aspects of our proposed GSFM framework. First, we compare the performance of GSFM models trained to predict $x_0$ directly versus those trained to predict the noise component, illustrating the advantages of our chosen prediction strategy. We, then, examine the impact of different sampling step lengths during inference on the quality and efficiency of the results. Additionally, we evaluate the computational and memory efficiency of our framework, highlighting its scalability for large-scale seismic datasets. Finally, we address the limitations of our approach and propose potential directions for future research to further enhance the capabilities of GSFM in seismic processing. 

\subsection{Comparison of predicting target and noise}
We here compare the performance of GSFM trained to predict target $x_0$ with a GSFM variant trained to predict noise $\epsilon$. Both models utilize the same network architecture and training configurations, but their target outputs differ. The noise-targeted GSFM aims to predict the noise component added during the forward diffusion process, while the $x_0$-targeted GSFM directly predicts the labeled data. We evaluate their respective performance, where we use denoising and interpolation tasks on synthetic data as an example. 

\begin{figure*}[htbp]
\centering
\includegraphics[width=\textwidth]{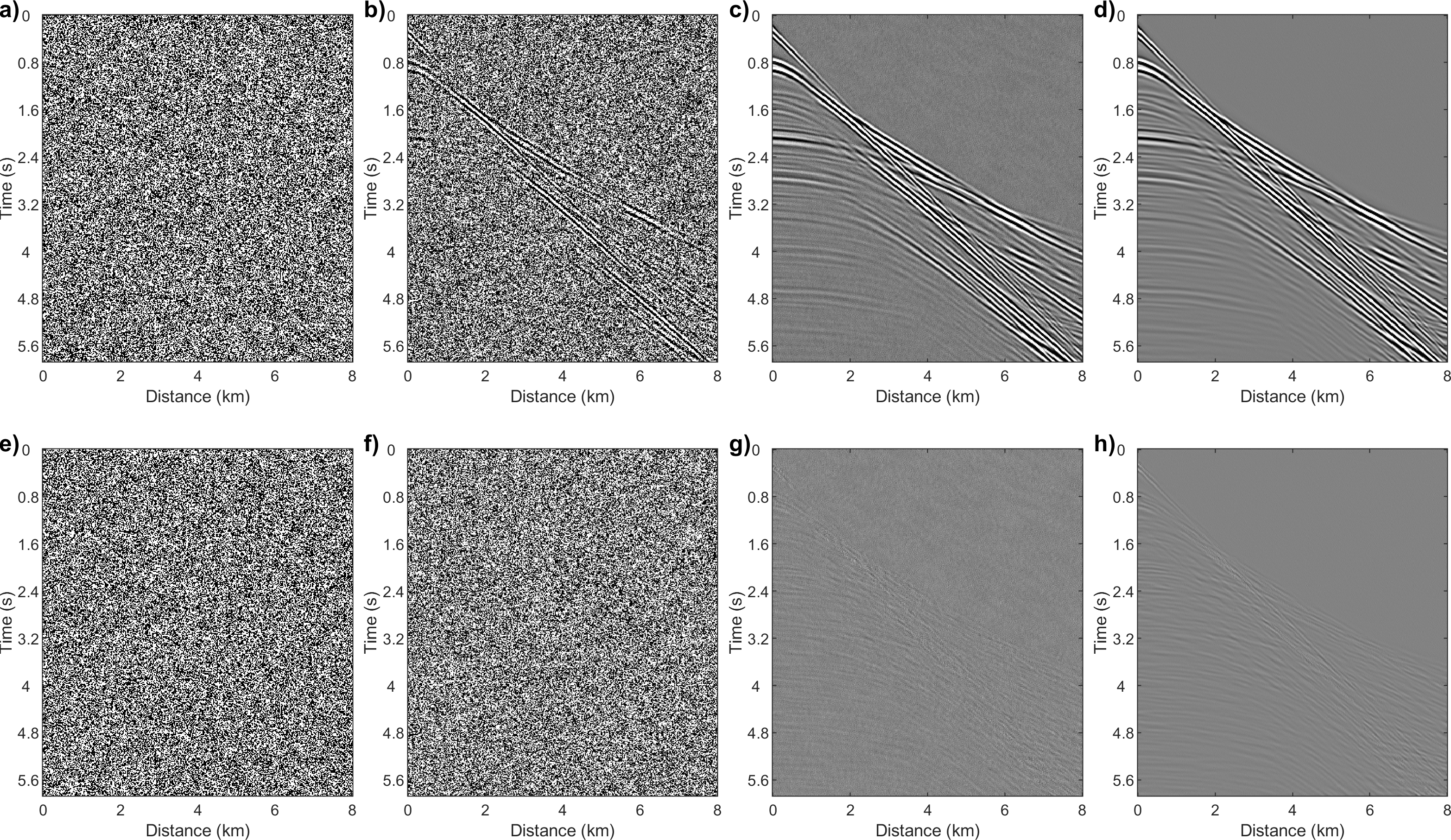}
\caption{Denoised products from the pre-trained DSFM when the prediction target is noise. The label and input test data can be found in Figures \ref{fig2}a and \ref{fig2}b, respectively. The denoised products using sampling step sizes (a) 1, (b) 100, (c) 500, and (d)
. (e, f, g, and h) The corresponding difference between the denoised results and the labeled data. }
\label{fig16}
\end{figure*}

\begin{figure*}[htbp]
\centering
\includegraphics[width=\textwidth]{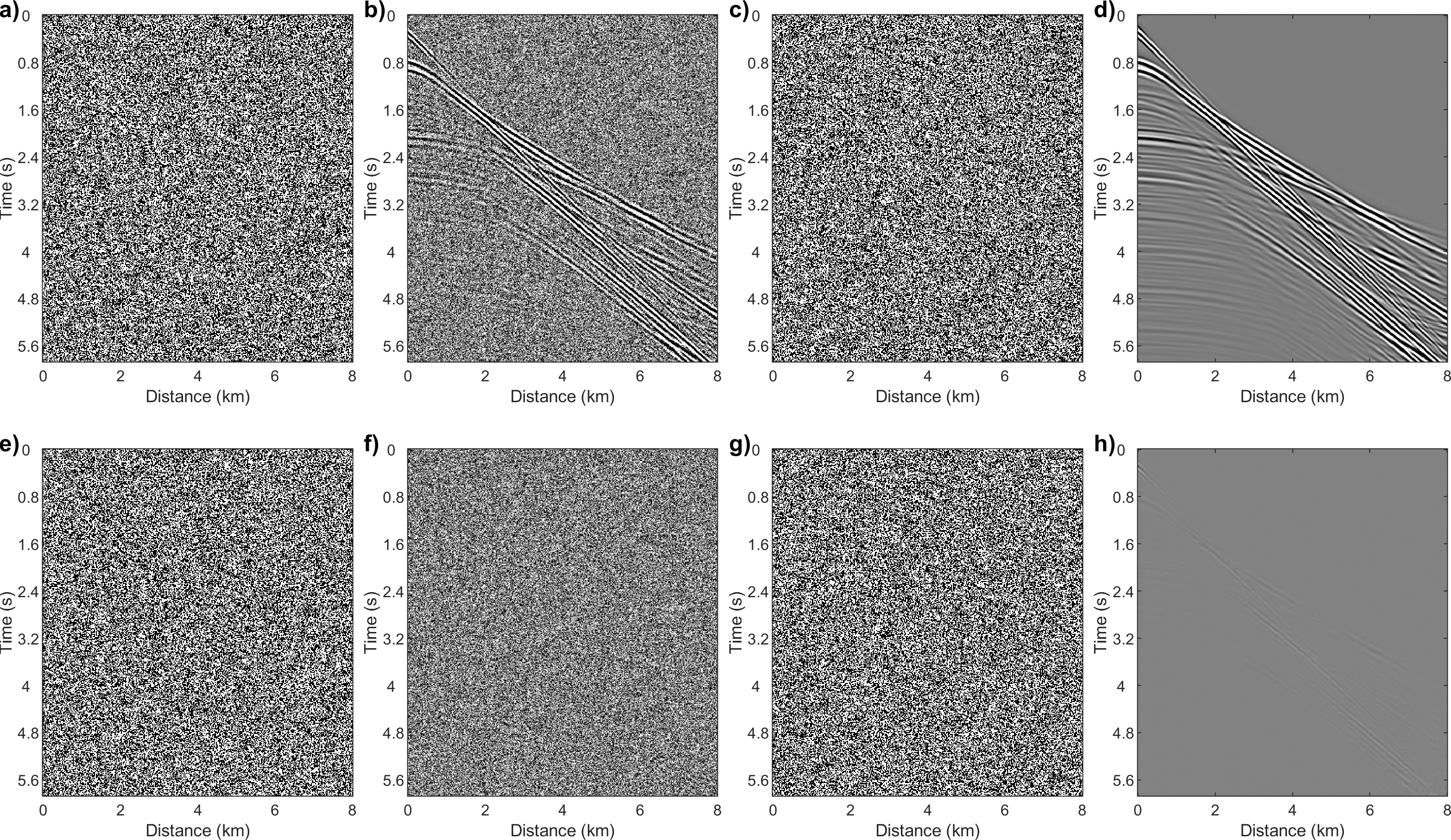}
\caption{Interpolation products from the pre-trained DSFM when the prediction target is noise. The label and input incomplete data can be found in Figures \ref{fig4}a and \ref{fig4}b, respectively. The interpolated products using sampling step sizes (a) 1, (b) 100, (c) 500, and (d) 1000. (e, f, g, and h) The corresponding difference between the interpolated results and the labeled data.}
\label{fig17}
\end{figure*}

Figures \ref{fig16} and \ref{fig17} illustrate the denoising and interpolation results generated by the noise-targeted pre-trained GSFM under different sampling step sizes, including $T=1$, 100, 500, and 1000. For the denoising task, the labeled and input data are shown in Figures \ref{fig2}a and \ref{fig2}b, respectively. Panels (a) to (d) in Figure \ref{fig16} display the denoised products obtained using the specified sampling step sizes, while panels (e) to (h) present the corresponding residuals compared to the labeled data. Similarly, for the interpolation task, the labeled and incomplete input data are shown in Figures \ref{fig4}a and \ref{fig4}b, respectively. Figure \ref{fig17} follows the same structure, highlighting the interpolation outputs and residuals. 

It is evident that the noise-targeted GSFM produces suboptimal results when using lower sampling step sizes ($T=1$, 100, and 500), with significant residual noise. This behavior highlights the challenges faced by the GSFM trained to predict noise, as it struggles to fully recover clean outputs within limited time steps. Even when the maximum sampling steps ($T=1000$) are used, noticeable signal leakage persist in the outputs. For instance, in the interpolation task, when comparing the results of the noise-targeted GSFM using $T=1000$ to the $x_0$-targeted GSFM using $T=1$ (see Figures \ref{fig4}c and \ref{fig4}f), the superiority of the latter becomes evident. While the $x_0$-targeted GSFM produces clean and coherent interpolated results with negligible residuals, the noise-targeted GSFM still contains significant signal leakage even with the highest sampling steps.  

Tables \ref{tab4} and \ref{tab5} further quantify these findings, where we calculate the MSE metric for denoising and interpolation tasks across varying noise and missing levels, respectively. The $x_0$-targeted GSFM consistently achieves lower MSE values than the noise-targeted GSFM, irrespective of the sampling step size. While the noise-targeted model shows general improvement with increased sampling steps, it requires the high step size ($T=500$ and 1000) to deliver results comparable to the $x_0$-targeted GSFM using step size $T=1$. Moreover, we can see that, the performance of the noise-targeted GSFM reveals further limitations. For certain noise levels (10\%, 20\%, 50\%, and 60\%) in the denoising task and missing levels (20\%) in the interpolation task, even when using the maximum sampling steps ($T=1000$), the resulting products still exhibit high MSE values. These elevated MSE scores, especially when compared to the consistently low MSE values achieved by the $x_0$-targeted GSFM, indicate that the noise-targeted GSFM has not been effectively optimized for these tasks. This lack of optimization becomes particularly issues in scenarios requiring fine-grained accuracy, as the noise-targeted GSFM struggles to converge to solutions that adequately approximate the labeled data.

\begin{table}[htbp]
    \centering
    \caption{Comparison of denoising MSE between $x_0$-targeted and noise-targeted GSFM across different noise levels}
    \label{tab4}
    \begin{tabular}{@{} c c c c c c @{}}
        \toprule
        \multirow{2}{*}{Noise Level} & \multirow{2}{*}{GSFM} & \multicolumn{4}{c}{Predicting noise} \\ \cmidrule(lr){3-6} 
         &  & $T=1$ & $T=100$ & $T=500$ & $T=1000$ \\ 
        \midrule
        10\% & \textbf{3.09e-07} & 1.57e-01 & 2.84e-02 & 1.01e-05 & 2.50e-01 \\
        20\% & \textbf{9.20e-07} & 1.63e-01 & 8.3e-03  & 1.28e-05 & 2.40e-01 \\
        30\% & \textbf{1.67e-06} & 1.64e-01 & 7.1e-03  & 1.58e-05 & 4.46e-06 \\
        40\% & \textbf{2.61e-06} & 1.61e-01 & 3.43e-05 & 1.46e-05 & 7.63e-06 \\
        50\% & \textbf{3.79e-06} & 1.61e-01 & 3.32e-05 & 1.73e-05 & 1.30e-01 \\
        60\% & \textbf{4.60e-06} & 1.60e-01 & 6.40e-03 & 4.50e-03 & 1.53e-01 \\
        \bottomrule
    \end{tabular}
\end{table}

\begin{table}[htbp]
    \centering
    \caption{Comparison of interpolation MSE between $x_0$-targeted and noise-targeted GSFM across different missing levels}
    \begin{tabular}{@{} c c c c c c @{}}
        \toprule
        \multirow{2}{*}{\centering Missing Level} & \multirow{2}{*}{\centering GSFM} & \multicolumn{4}{c}{Predicting noise} \\ \cmidrule(lr){3-6}
         &  & $T=1$ & $T=100$ & $T=500$ & $T=1000$ \\ 
        \hline
        10\% & \textbf{1.45e-08} & 1.58e-01 & 9.20e-03 & 1.28e-05 & 3.88e-06 \\
        20\% & \textbf{1.93e-08} & 1.66e-01 & 9.63e-04 & 1.27e-05 & 2.50e-01 \\
        30\% & \textbf{2.85e-08} & 1.66e-01 & 1.90e-05 & 4.93e-05 & 3.03e-06 \\
        40\% & \textbf{8.54e-08} & 1.62e-01 & 3.20e-05 & 1.13e-05 & 2.39e-06 \\
        50\% & \textbf{4.08e-08} & 1.65e-01 & 1.10e-03 & 2.17e-01 & 1.80e-06 \\
        60\% & \textbf{3.65e-07} & 1.60e-01 & 4.40e-03 & 1.32e-05 & 3.65e-06 \\
        \bottomrule
    \end{tabular}
    \label{tab5}
\end{table}

\subsection{Comparison of different sampling steps}
As stated earlier, the GDMs begin with random noise and gradually reduce the noise through a sequence of iterative steps to generate the final prediction. In the previous subsection, we observed that the performance of the GSFM trained to predict noise generally improves with an increasing number of sampling steps. However, for our GSFM designed to predict $x_0$, the synthetic and field data examples consistently used only one sampling step to maintain alignment with the benchmark methods and, also, to consider computational efficiency. This raises an important question: does using more sampling steps lead to better prediction quality for our $x_0$-targeted GSFM? 

To investigate this, we test the $x_0$-targeted GSFM with different sampling step configurations ($T=1$, 10, 50, 100, 500,and 1000) using synthetic test data. The corresponding MSE metrics for all four tasks are summarized in Table \ref{tab6}. We can see that the performance of our $x_0$-targeted GSFM remains highly stable across varying sampling steps. This finding suggests that our $x_0$-targeted GSFM is highly efficient and robust, achieving optimal performance with minimal sampling steps. This is an important advantage for practical applications, as it reduces the computational burden associated with the iterative nature of GDMs. Unlike noise-targeted models, where longer sampling steps are often necessary to achieve acceptable results, the $x_0$-targeted GSFM demonstrates consistent performance even with a single sampling step, highlighting its practical value in SPTs. 

\begin{table}[htbp]
    \centering
    \caption{The MSE metric of our \(x_0\)-targeted GSFM using different sampling steps}
    \label{tab6}
    \begin{tabular}{@{} c c c c c c c @{}}
        \toprule
        Task & \(T=1\) & \(T=10\) & \(T=50\) & \(T=100\) & \(T=500\) & \(T=1000\) \\ 
        \midrule
        Denoising                        & 1.67e-06 & 1.61e-06 & 1.61e-06 & 1.61e-06 & 1.61e-06 & 1.61e-06 \\
        Backscattered noise attenuation & 9.59e-07 & 9.59e-07 & 9.59e-07 & 9.59e-07 & 9.59e-07 & 9.58e-07 \\
        Interpolation                   & 4.08e-08 & 4.08e-08 & 4.07e-08 & 4.08e-08 & 4.07e-08 & 4.07e-08 \\
        Low-frequency extrapolation     & 6.0e-07  & 5.91e-07 & 6.08e-07 & 6.16e-07 & 6.05e-07 & 6.05e-07 \\
        \bottomrule
    \end{tabular}
\end{table}

\subsection{Computation and memory consumption}
Our GSFM employs an U-Net-based network architecture, which is shared with the two benchmarks provided in this study, differing only in minor aspects. Consequently, the computational time and memory consumption for pre-training our GSFM are nearly identical to those of the benchmarks. This consistency extends to the inference phase as well, where the GSFM demonstrates remarkable efficiency. 

During the fine-tuning stage, the iterative refinement strategy allows for progressive improvement of the GSFM's performance on real data. The time consumed during fine-tuning depends on the decision to terminate refinement based on the model's output at the current stage. In our field data examples, we observed that using the same total number of iterations as the conventional pre-training fine-tuning paradigm already produced satisfactory results with the iterative refinement strategy. This demonstrates that the cost of fine-tuning GSFM remains practical and manageable. 

By adopting an $x_0$-target-based GDM, our GSFM achieves comparable results with a single sampling step to those obtained with multiple sampling steps. This unique feature significantly reduces the computational burden during inference stage. As a result, the time and memory requirements for inference in our GSFM are also comparable to those of the benchmarks. 

Overall, the computational and memory efficiency of our GSFM is comparable to the benchmarks during both pre-training, fine-tuning, and also inference stages. Furthermore, the iterative fine-tuning strategy ensures that the additional time cost associated with refinement remains reasonable, making GSFM a computationally efficient solution for SPTs. 

\subsection{Limitations and Future work}
The core idea behind our GSFM is to leverage GDMs to capture and learn the joint distribution of seismic data. Specifically, we consider seismic shot gathers and define the "perfect" shot gather as one that is clean, complete, and broadband (also for example in the future free of multiples). By pre-training the GSFM on synthetic data and fine-tuning it on field data, we aim for GSFM to capture these desirable distribution characteristics. However, certain SPTs require transformations across fundamentally different domains, which present unique challenges. 

For example, tasks like velocity analysis require transforming seismic shot gathers into a smooth background velocity model, effectively converting data from the time domain to the depth domain. To adapt GSFM for such tasks, we could retain the shot gather as one input channel while providing the target background velocity model as the label in another channel. In this case, the GSFM would need to learn the distribution of background velocity models. However, the distribution of background velocity models is inherently disjoint from the distribution of seismic shot gathers. This divergence poses significant challenges for GSFM optimization, as it becomes difficult for the model to learn and represent such disparate distributions within the same framework. 

This limitation highlights a current constraint of our GSFM: it struggles to handle tasks where the target data is not of the same domain as the input data. For pure seismic preprocesing, this is fine. However, for additional tasks like velocity model building, this is a limitation. Addressing this limitation is a key focus for future research. In subsequent work, we aim to extend the capabilities of GSFM by developing strategies to effectively handle such non-overlapping distributions. This may involve introducing modular designs or hybrid frameworks that allow the model to adaptively learn and represent multiple distinct distributions, enabling broader applicability to a wider range of SPTs.

\section{Conclusions}
We introduced the generative seismic foundation model (GSFM), a novel framework built upon generative diffusion models (GDMs) to address multi-task seismic processing. GSFM leverages a generative approach to learn and capture the underlying joint distribution of seismic data, aiming to represent clean, complete, and broadband characteristics. By encoding tasks with class labels and integrating synthetic pre-training with iterative fine-tuning on field data, GSFM achieves a unified framework for seismic denoising, backscattered noise attenuation, interpolation, and low-frequency extrapolation. 

On synthetic data, our pre-trained GSFM achieved performance comparable to traditional pre-training strategies followed by extensive fine-tuning, while significantly outperforming a benchmark model with the same architecture across all tasks. The results demonstrated that GSFM, even without task-specific adjustments, delivers robust and high-quality processing outputs. On field data, the iterative fine-tuning strategy we proposed effectively addressed the generalization challenges inherent in traditional pre-training and fine-tuning paradigms. The fine-tuned GSFM consistently outperformed both benchmarks, establishing its ability to adapt to field data distributions while preserving computational efficiency. 

Through comparative experiments, we demonstrated that our iterative fine-tuning strategy is optimal for refining GSFM on field data. This strategy not only improved processing performance but also provided a clear guideline for applying our pre-trained GSFM to real-world scenarios. Furthermore, the uncertainty quantification capability of our GSFM highlighted its potential for evaluating the reliability of processing results, adding a layer of interpretability that is critical for decision-making in seismic workflows. Also, we can use uncertainty quantification to gudie our fine-tuning stage and, thus, to evaluate whether our GSFM is properly trained. 

In summary, GSFM represents a significant step forward in seismic processing by unifying multiple tasks under a single generative framework. Its ability to generalize across synthetic and field data, coupled with its efficiency and versatility, demonstrates the value of incorporating GDMs into geophysical applications. While challenges remain, such as addressing different input-target distributions, GSFM establishes a strong foundation for future research, offering practical solutions for seismic processing.
\section*{Acknowledgments}
This publication is based on work supported by the King Abdullah University of Science and Technology (KAUST). The authors thank the DeepWave sponsors for supporting this research. This work utilized the resources of the Supercomputing Laboratory at King Abdullah University of Science and Technology (KAUST) in Thuwal, Saudi Arabia.
\section*{Code and Data Availability}
The data and accompanying codes that support the findings of this study are available at: 
\url{https://github.com/DeepWave-KAUST/GSFM}. (During the review process, the repository is private. Once the manuscript is accepted, we will make it public.)

\bibliographystyle{unsrtnat}
\bibliography{references}

\end{document}